\documentclass[superscriptaddress,secnumarabic,amssymb,amsmath,nobibnotes,aps,prd,nofootinbib,onecolumn]{revtex4}
\pdfoutput=1
\usepackage{graphicx}
\usepackage{epstopdf}
\usepackage{epsf}
\usepackage{bm}
\usepackage{amsmath,mathtools,commath}
\usepackage{amsfonts}
\usepackage{amssymb}%

\usepackage{xcolor}
\usepackage{hyperref}
\usepackage{physics}

\usepackage{pgfplots}
\pgfplotsset{compat=newest}
\usepackage{placeins}

\providecommand{\U}[1]{\protect\rule{.1in}{.1in}}

\newcommand{\be}{\begin{equation}}
\newcommand{\ee}{\end{equation}}

\newcommand{\mincir}{\raise
-3.truept\hbox{\rlap{\hbox{$\sim$}}\raise4.truept\hbox{$<$}\ }}
\newcommand{\magcir}{\raise
-3.truept\hbox{\rlap{\hbox{$\sim$}}\raise4.truept\hbox{$>$}\ }}

\newcommand{\ud}{\,\mathrm{d}}
\newcommand{\uD}{\,\text{D}}
\newcommand{\Mpl}{M_{\mathrm{Pl}}}
\usepackage{comment}

\begin{document}

\title{(Slow-)Twisting inflationary attractors}

\author{Perseas Christodoulidis}
\email{perseas@ewha.ac.kr}
\affiliation{Department of Science Education, Ewha Womans University, Seoul 03760, Republic of Korea}

\author{Robert Rosati}
\email{robert.j.rosati@nasa.gov}
\affiliation{NASA Postdoctoral Program Fellow, NASA Marshall Space Flight Center, Huntsville, AL 35812, USA}

\begin{abstract}
We explore in detail the dynamics of multi-field inflationary models. We first revisit the two-field case and rederive the coordinate independent expression for the attractor solution with either small or large turn rate, emphasizing the role of isometries for the existence of rapid-turn solutions.
Then, for three fields in the 
\textit{slow-twist} 
regime we provide elegant expressions for the attractor solution for generic field-space geometries and potentials and study the behaviour of first order perturbations. For generic $\mathcal{N}$-field models, our method quickly grows in algebraic complexity. We observe that field-space isometries are common in the literature and are able to obtain the attractor solutions and deduce stability for some isometry classes of $\mathcal{N}$-field models. Finally, we apply our discussion to concrete supergravity models. These analyses conclusively demonstrate the existence of $\mathcal{N}>2$ dynamical attractors distinct from the two-field case, and provide tools useful for future studies of their phenomenology in the cosmic microwave background and stochastic gravitational wave spectrum.
\end{abstract}

\maketitle

\tableofcontents

\section{Introduction}

The theory of cosmological inflation, originally introduced as a scenario that evades the fine-tunings of the classical FLRW universe \cite{Starobinsky:1980te,Sato:1981ds,Sato:1980yn,Kazanas:1980tx,Guth:1980zm,Linde:1981mu,Albrecht:1982wi}, is currently the leading paradigm for the origin of structure formation \cite{Akrami:2018odb}. In its simplest form, the evolution of a scalar field over flat regions of its potential causes the accelerated expansion of the universe. One might think that inflation is also subject to fine-tuning problems of its own, because the onset of the accelerating expansion is not guaranteed for all initial conditions. This turns out not to be the case due to the dissipative nature of the cosmological equations in an expanding background which yields a notion of initial conditions independence for models with a single scalar field. This fact makes single-field models of inflation both self-consistent and phenomenologically successful in matching the CMB observations to a great accuracy. On a theoretical level this success is eclipsed by obstacles in embedding the inflationary paradigm into high-energy physics. 

More specifically, high-energy theories, such as string theory or supergravity, in general require consideration of the dynamics of multiple scalar fields in the early universe, and through the swampland program, potentially limit their allowed interactions (e.g. \cite{Obied:2018sgi,Ooguri:2018wrx,Garg:2018reu,Garg:2018zdg,Denef:2018etk,Andriot:2018mav,Achucarro:2018vey,Scalisi:2018eaz,Palti:2019pca}).
For a multi-field model to be consistent with the general philosophy of inflation, observables should have a weak dependence on initial field configurations; this can only be achieved if any additional degrees of freedom quickly become non-dynamical leaving just one field to drive the evolution.\footnote{For the simplest multi-field models without any hierarchies between the model's parameters (such as the masses of the fields or parameters that quantify the strength of the field space curvature) the non-uniqueness of observables is quite evident at the two-field level, (see e.g.~\cite{Frazer:2013zoa}), and, moreover, it persists even in the many-field limit \cite{Christodoulidis:2019hhq}.} The truncation to one field is tricky and often times the dynamics of the effective description is quite distinct from what one obtains from purely single-field models at the level of both the background and the perturbations (see e.g.~\cite{Tolley:2009fg,Brown:2017osf,Mizuno:2017idt,Christodoulidis:2018qdw,Garcia-Saenz:2018ifx,Fumagalli:2019noh,Bjorkmo:2019aev,Aragam:2019omo,Garcia-Saenz:2019njm,Chakraborty:2019dfh,Dimakis:2019qfs,Paliathanasis:2020wjl,Bounakis:2020xaw}). In certain cases, even though the background dynamics can be reduced to the dynamics of a single-field, the behaviour of perturbations can be drastically different; the effect of isocurvature perturbations can be absorbed into the definition of a nontrivial speed of sound in the evolution of the curvature perturbation, leading to deviations from the single-field predictions (see for instance \cite{Achucarro:2010da,Achucarro:2012sm,Cespedes:2012hu,Achucarro:2012yr,Gwyn:2012mw,Achucarro:2016fby,Wang:2019gok}). 

Focusing on the background, it becomes apparent that consistent multi-field models should display a strong attractor behaviour, akin to single-field models, that weakens their initial conditions dependence. The extra fields should remain non-dynamical during the relevant evolution (at least 50-60 $e$-folds before the end of inflation) and, thus, the system follows a specific trajectory in the field space that we call the attractor solution. In general, the attractor solution is not equivalent to the solution of a single-field model because of the existence of turns in the trajectory; this happens when the extra fields remain non-dynamical but are excited away from their respective minima of the potential. This generic multi-field dynamics at the background level has been explored recently in Ref.~\cite{Bjorkmo:2019fls,Aragam:2020uqi}, showing the existence of rapid-turn solutions based on a generic calculation of the turn rate, and in Ref.~\cite{Christodoulidis:2019mkj} where analogous expressions for the late-time solution were provided using an effective potential that dictates dynamics at late-times. Moreover, this class of models were shown to lie at the intersection of de-Sitter and scaling solutions where every ``slow-roll''-type approximation becomes exact \cite{Christodoulidis:2019jsx}. It remains unclear whether two-field studies capture all essential features of the multi- or many-field evolution, especially for concrete models derived from supergravity.
Many studies of $\mathcal{N}>2$ inflation exist in the literature, including \cite{Dimopoulos:2005ac, Piao:2006nm, Kaiser:2012ak, McAllister:2012am, Cespedes:2013rda, Easther:2013rva, Price:2014ufa, Dias:2016slx, Dias:2017gva, Paban:2018ole, Aragam:2019omo, Pinol:2020kvw, Christodoulidis:2021vye}. Many of these works have focused on specific types of slow-turn inflationary models, the universality of the many-field limit, or the effective field theory of the curvature perturbation for $\mathcal{N}>2$ models, overall leaving generic multi-field background trajectories unexplored.

In this work we study the model-independent attractor solutions of multi-field inflation, providing analytic expressions for three-field models.\footnote{During the preparation of this paper the preprint \cite{2210.00031} was uploaded to the arXiv, which has as one of their conclusions that slow-roll, rapid-turn behavior is short-lived or very rare in two-field inflation. This claim seems to be in tension with previous literature, where rapid-turn models can easily be constructed for highly curved spaces, and further work is necessary to reconcile them.} The expressions are particularly tractable in the slow-twist and rapid-turn regime. We additionally explore the inflationary perturbations of these three-field solutions, and recover an effective single-field description. We also generalize our solutions to four or more fields by assuming field spaces with a sufficiently high number of isometries. The paper is organized as follows: in Sec.~\ref{sec:frenet_serret} we derive the evolution equations for the first three basis vectors of the orthonormal Frenet-Serret system and calculate the set of higher order bending parameters that parameterize how the field-space trajectory bends in the $\mathcal{N}$-dimensional space. In Sec.~\ref{sec:two_field} we perform a complete study of two-field models under the assumption that the equations of motion admit slow-roll-like solutions. Next, in Sec.~\ref{sec:3field} we argue that the complexity of the three-field problem makes it impossible to adopt a similar generic treatment and which forces us to consider certain simplifications, such as small torsion. In Sec.~\ref{sec:isometries} we focus on specific $\mathcal{N}$-field problems with isometries and discuss rapid-turn solutions and their stability. We analyse in more detail observables for three-field models in Sec.~\ref{sec:observables_stability}. Finally, in Sec.~\ref{sec:summary} we offer our conclusions.
\\

\textbf{Conventions:} We will use $N$ for the $e$-folding number, defined from $\ud N = H \ud t$, and $\mathcal{N}$ for the number of fields; $G_{ij}$ denotes the field-space metric; $t^i, n^i,b^i$ represent the components of the first three basis vectors of the orthonormal Frenet system. Since we will use multiple different bases, to suppress notation we will use lower case Greek letters ($\alpha,\beta,\gamma,\cdots$) to denote indices belonging to the orthonormal kinematic basis while middle lower case Latin indices ($i,j,k,\cdots$) represent general field metric indices. We work in units with $\Mpl^2=1$.

\section{Multi-field Trajectories} \label{sec:frenet_serret}
In this section we set the foundations for the rest of this work, and study inflationary trajectories in $\mathcal{N} > 2$ field space and compute the bending parameters as generically as possible.
Our goal is to express these kinematic quantities in terms of the field space metric and potential using the equations of motion.

We study the evolution of $\mathcal{N}$ scalar fields minimally coupled to gravity
\begin{align}
    S = \int \dd[4]{x} \sqrt{-g} \left[ \frac{1}{2} R_{e} -\frac{1}{2}G_{ij}(\phi^k)\partial_\mu \phi^i \partial^\mu \phi^j - V(\phi^k) \right]
    \label{eq:introlagrangian}
\end{align}
where $R_e$ is the Ricci scalar associated with the spacetime metric and the fields interact via the field-space metric $G_{ij}(\phi^k)$ and the potential $V(\phi^k)$. 

As standard in the inflationary literature, we consider the fields spatially homogeneous at the classical level.
The classical equations of motion can be written
\begin{align}
\uD_N (\phi^i)^\prime + (3-\epsilon) (\phi^i)^\prime + \frac{G^{ij}V_{,j}}{H^2} = 0,
\label{eq:eomFieldBasis}
\end{align}
where primes denote e-fold derivatives $\ud N = H \ud t$, $\uD_N (\phi^i)^\prime \equiv (\phi^j)^\prime \uD_j (\phi^i)^\prime = (\phi^i)''+ \Gamma^i_{jk} (\phi^j)^\prime(\phi^k)^\prime$, and the $\Gamma^i_{jk}$ are the connection components associated to the field space metric $G_{ij}$.

We define the slow-roll parameters
\begin{align}
\epsilon &\equiv \frac{1}{2} (\phi^i)^\prime G_{ij} (\phi^j)^\prime,\\
\eta &\equiv \epsilon^\prime / \epsilon \, ,
\end{align}
which probe the kinetic energy and acceleration of the fields respectively.

\begin{figure}
\centering
\includegraphics[width=0.4 \textwidth]{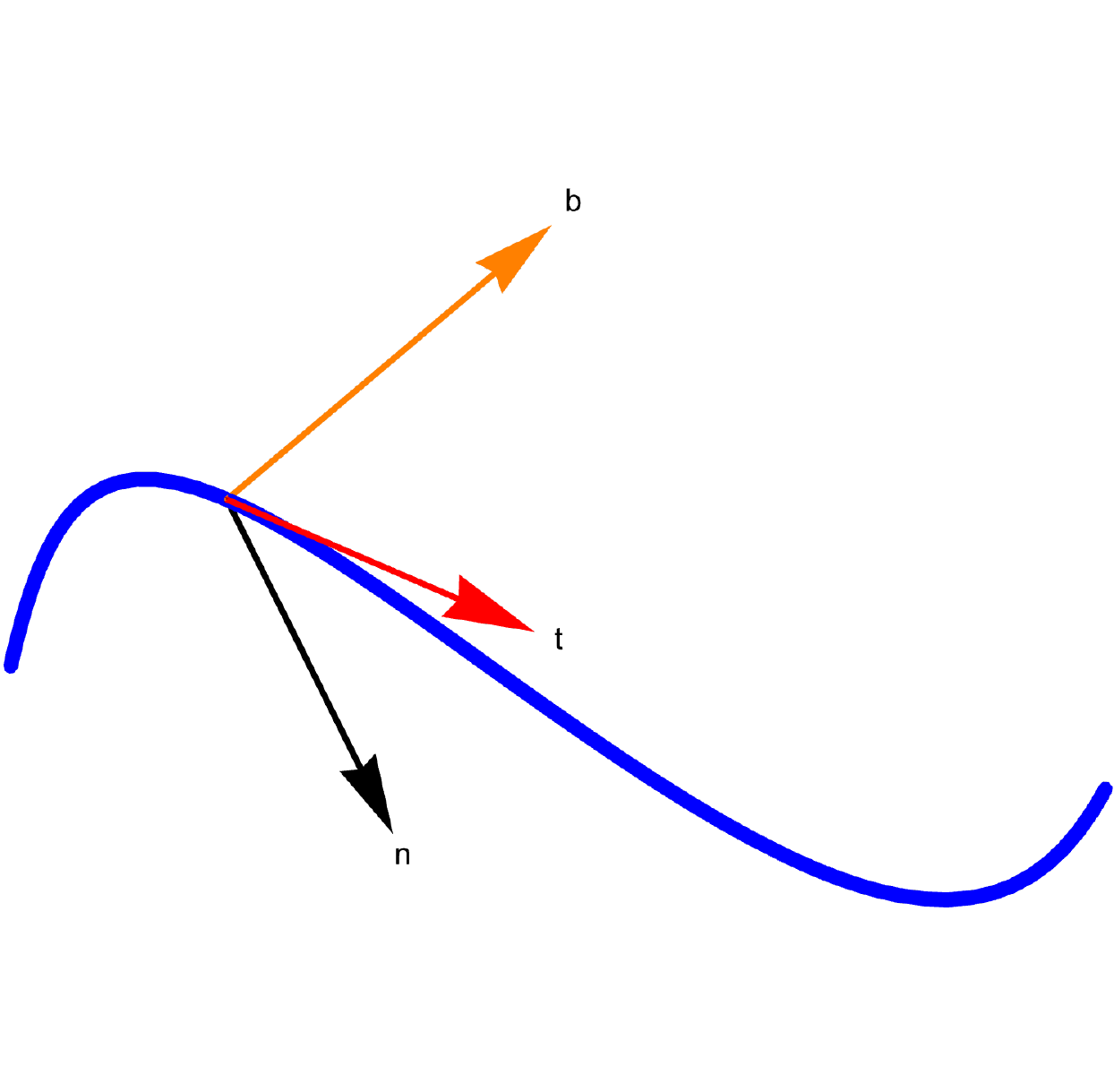}
\begin{tikzpicture}
    \begin{axis}[
    		title={$0 < T < \Omega \, ,~T = \mathcal{O}( \Omega)$},
            axis lines=left,
            width=220bp,
            height=220bp,
            xmin=-1.5,xmax=1.5,
            ymin=-1.5,ymax=1.5,
            zmin=-2,zmax=2,
            xlabel={$\phi_1$},
            ylabel={$\phi_2$},
            zlabel={$\phi_3$},
            view={60}{30},
            xtick = \empty,
            ytick = \empty,
            ztick = \empty,
            smooth
            ]
        \addplot3[
        		ultra thick,
        		samples y=0,
        		mesh,
                samples=100,
                domain=-1:1,
                ]
                ({cos(deg(2*pi*x))},{sin(deg(2*pi*x))},{x});
    \end{axis}
\end{tikzpicture}
\caption{(Left) Pictorial representation of the orthonormal kinematic system on a three-dimensional curve. (Right) The helix is the prototypical example of a curve with constant turn rate and torsion.}
\label{fig:frenetSerretSketch}
\end{figure}
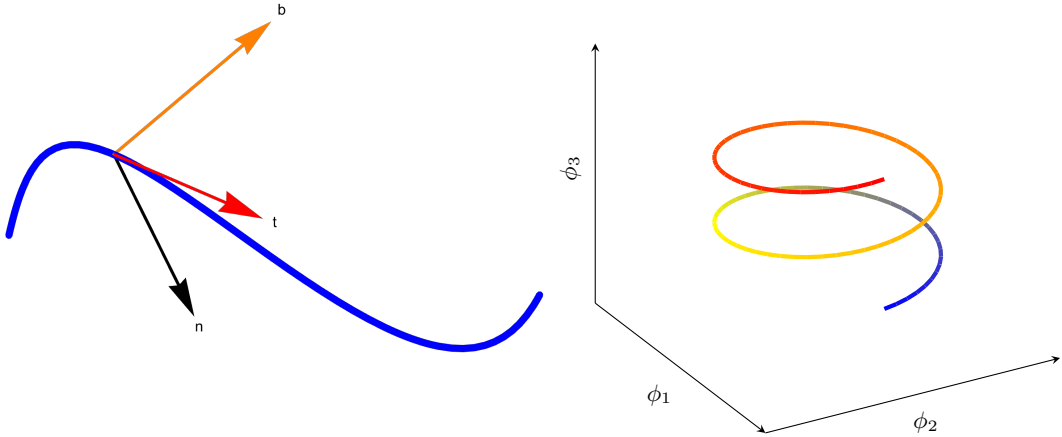

We are interested in the turning parameters of the trajectory, which are defined in the kinetic orthonormal basis derived from the fields' motion. This basis is formed from the velocity unit vector $t^i \equiv (\phi^i)^\prime / \sqrt{2\epsilon}$ and its derivatives.
The change of the velocity unit vector defines the turn rate of the inflationary trajectory, as well as the normal vector to the trajectory
\begin{align}
\uD_N t^i \equiv \Omega n^i.
\end{align}
Similarly, the derivative of the normal vector defines the torsion or twist rate of the trajectory, as well as the binormal vector
\begin{align}
\uD_N n^i\equiv -\Omega t^i + T b^i.
\end{align}
Further derivatives define higher-order bending parameters, for $\mathcal{N}-1$ total.
This kinematic basis can be neatly summarized in the Frenet-Serret system of equations
\cite{kreyszig2013differential}
\begin{equation}
\uD_{N} 
\begin{pmatrix}
t^i \\
n^i \\
b^i \\
b_2^i \\
\vdots 
\end{pmatrix} 
=
\begin{pmatrix}
0 & \Omega & 0  & 0 &\cdots \\
-\Omega & 0 & T &  0 & \cdots \\
0 &  -T &0 &  T_2 & \cdots \\
0 &  0 & - T_2 &  0 & \cdots \\
\vdots  &\vdots   & \vdots & \vdots  & \ddots  
\end{pmatrix} 
\begin{pmatrix}
t^i \\
n^i \\
b^i \\
b_2^i \\
\vdots  
\end{pmatrix} 
\, ,
\label{eq:FrenetSerretEfolds}
\end{equation}
where $b_j^i, T_j$ with $j \geq 2$ define additional bending parameters when $\mathcal{N} > 3$ (see Fig.~\ref{fig:frenetSerretSketch} for examples).

The equations of motion are particularly simple in this kinetic basis, where \eqref{eq:eomFieldBasis} becomes
\begin{align}
\Omega n^i + \left(3-\epsilon+\frac{\eta}{2}\right) t^i + \frac{G^{ij}V_{,j}}{\sqrt{2 \epsilon} H^2} = 0.
\label{eq:eomKinetic}
\end{align}
From here we can read that (on-shell) the potential gradient only has components along $t^i$ and $n^i$, and also that the turn rate $\Omega$ can be expressed
\begin{align}
\Omega n^i \equiv \uD_N t^i = \frac{3-\epsilon}{\sqrt{2\epsilon}} \left( w_\sigma t^i - w^i \right)
\label{eq:OmegaVector}
\end{align}
where we used the Friedmann equation $V = H^2 (3-\epsilon)$ and defined $w_i \equiv V_{,i} / V$ and $w_\sigma \equiv w^i t_i$.

For convenience, below we work with the trajectory's arclength parameter $\sigma$, defined so that $\uD_N \equiv \sqrt{2\epsilon} \uD_\sigma$.\footnote{It is worth mentioning that the use of $\sigma$ as the independent ``time'' parameter is also physical relevant.
Often the turning parameters in \eqref{eq:FrenetSerretEfolds} and $\epsilon$ are analytically related, as in Sec.~\ref{subsec:diagonal_isometry} and \cite{Aragam:2019omo,Christodoulidis:2018qdw}.
In models without a known analytic relationship, we often find numerically that for either slow- or rapid-turn models the quantities that remain almost constant are the bending parameters $k$ and $\tau$.}
The equivalent bending parameters are defined as $k \equiv \Omega / \sqrt{2\epsilon}$, $\tau \equiv T/\sqrt{2\epsilon}$, $\tau_j \equiv T_j / \sqrt{2\epsilon}$, etc.
Through additional derivatives of \eqref{eq:OmegaVector}, we can find expressions for the bending parameters in terms of only kinematic quantities and covariant derivatives of the potential.
We leave the details to Appendix \ref{sec:frenet_serret_app}, but quote the torsion vector here 
\begin{equation} \label{eq:torsion}
\tau b^i =\left[ \left(\ln{3-\epsilon \over 2\epsilon k} \right)_{,\sigma } +  {3 - \epsilon \over 2\epsilon}  w_{\sigma}  \right] n^i  + {3 - \epsilon \over 2\epsilon k} w_{\sigma \sigma}t^i    -  {3 - \epsilon \over 2 \epsilon k}  w^{i}_\sigma  \, .
\end{equation}

From these and similar expressions, it is possible to read off kinematic relationships for many of the potential derivatives, which we leave to the appendices but use when applicable.

Below we describe the late-time solutions in terms of these kinematic quantities, assuming only slow-roll.

\section{Two-field solutions} \label{sec:two_field}
\subsection{Coordinate-independent expression for the attractor solution}
In this section we present a neat way to find the generic slow-roll two-field attractor solution, before exploring higher dimensional field spaces. 
The solution is presented in a manifestly covariant expression when expressed in terms of geometric objects constructed by the field metric and the potential. The simplest objects we can construct include the trace of products of the Hessian $w_{ij}$ and projections of products of the Hessian along the gradient directions:
\begin{align} \label{eq:cn}
c_n \equiv  w^{i} \underbrace{w_{~i}^{j} \cdots w_{~m}^{l}}_\text{n times}  w^{m}  \, , \\ \label{eq:dn}
d_n \equiv \text{Tr} \left(\underbrace{w_{~i}^{j} \cdots w_{~m}^{l}}_\text{n times} \right) \, .
\end{align}
With these objects to our disposal, the next step is to use an appropriate basis that relates the right hand side of \eqref{eq:cn}, \eqref{eq:dn} with the slow-roll parameter $\epsilon$ and the turn rate and then form a system with a sufficient number of these curvature invariants that allows us to solve back for $\epsilon$.\footnote{One could also consider contractions with objects such as $V_{;ijk}$ containing three or more covariant derivatives. However, in this case the Frenet-Serret equations will constrain only a small number of them that inculde at least one $\sigma$ component, leaving a larger number of unknown parameters and hence a larger number of equations that have to be solved simultaneously for $\epsilon$.} We find it easier to work with the usual kinematic frame in a manner similar to Ref.~\cite{Christodoulidis:2019mkj} which, however, made use of a special coordinate system. We also avoid the potential gradient-based orthonormal basis used in Refs.~\cite{Bjorkmo:2019fls,Aragam:2020uqi}.

We assume the attractor solution has a slowly-changing $\epsilon$, so that $\eta$ is negligible in the equations of motion \eqref{eq:eomKinetic}. This assumption allows us to reduce the second order differential equation to an algebraic equation for the velocity, which implies some sort of a late-time solution. From the equations of motion we find the adiabatic component of the gradient vector as 
of the equations of motion as
\begin{equation} \label{eq:attractor_gradient}
w_\sigma = -\sqrt{2\epsilon} \frac{3-\epsilon + \eta/2}{3-\epsilon} \, ,
\end{equation}
and plugging this expression back into the definition of the turn rate we obtain 
\begin{equation}\label{eq:omega_def}
{1 \over 2} w^i w_i \equiv \epsilon_V = \epsilon \left[  \left( 1 + {\eta \over 2(3-\epsilon)} \right)^2  + \left({\Omega \over 3-\epsilon} \right)^2 \right] \, .
\end{equation}
This is our first expression that relates $\epsilon$ with the turn rate and the norm of the gradient vector. When the slow-roll conditions are satisfied ($\epsilon,|\eta|\ll 1$) then Eq.~\eqref{eq:omega_def} reduces to \cite{Hetz:2016ics}
\begin{equation} \label{eq:eps_epsV}
\epsilon_V \approx \epsilon \left(1 + {1 \over 9} \Omega^2 \right) \, ,
\end{equation}
while $\epsilon$ is approximated by 
\begin{equation} \label{eq:asymp_epsilon}
\epsilon \approx \epsilon_{\rm ad} \equiv {1  \over 2} \left({V_{\sigma} \over V } \right)^2 = {1  \over 2} w_{\sigma}^2 \, .
\end{equation}

To proceed we need an expression for the turn rate and we will look at the components of the Hessian. Projecting Eq.~\eqref{eq:torsion} along the normal vector relates the $_{n \sigma}$ component with the turn rate as
\begin{equation}
 w_{n \sigma }  = \left({3-\epsilon \over 2\epsilon k} \right)_{,\sigma } +  k w_{\sigma}  \, .
\end{equation}
Moreover, the $_{\sigma\sigma}$ component of the Hessian is also directly related to the turn rate; taking the derivative of Eq.~\eqref{eq:OmegaVector} we find
\begin{equation} \label{eq:wsigmasigma}
w_{\sigma \sigma} = { \Omega^2 \over 3- \epsilon}  - {1 \over 2} \eta - {\eta' \over 2(3 - \epsilon)} -  {\eta^2 \over 4(3 - \epsilon)} -  {\epsilon \eta^2 \over 2(3 - \epsilon)^2} \, .
\end{equation}
Finally, the $_{nn}$ component of the Hessian is not related to kinematic quantities and will be left as a free parameter.

Neglecting all $N$-derivatives of kinematic quantities, the Hessian in the kinematic frame takes the following form
\begin{equation}
w_{\alpha \beta} = 
\begin{pmatrix}
w_{\sigma \sigma} & w_{n \sigma} \\
w_{n \sigma} & w_{n n} 
\end{pmatrix} 
\approx
\begin{pmatrix}
{\Omega^2 \over 3- \epsilon} & - \Omega \\
 - \Omega & w_{n n} 
\end{pmatrix} \, .
\end{equation}
The Hessian contains three unknown quantities so we need two curvature invariants in addition to $\epsilon_V$. Choosing $c_1$ and $c_2$ given in the kinematic basis as
\begin{align}
c_1 = w_{\alpha \beta}w_{\alpha} w_{\beta}  \, , \qquad c_2 = w_{\alpha \beta} w_{\beta \gamma}w_{\alpha} w_{\gamma} \, ,
\end{align}
allows us to find a particularly simple expression for $\epsilon$, whenever $c_1,c_2\neq 0$
\begin{equation} \label{eq:epsilon1}
\epsilon = \epsilon_V - {c_1^2 \over 2 c_2} \, .
\end{equation}
Since $c_2 \geq 0$ we have strictly $\epsilon \leq \epsilon_V$ and this is a compatible with Eq.~\eqref{eq:omega_def} for $\eta=0$. In case $c_2=0$ (and hence also $c_1=0$) we can use the trace of the Hessian as an alternative curvature invariant
\begin{equation}
d_1 = \text{Tr}(w) =  G^{ij} w_{i;j} \, ,
\end{equation}
and find the following expression for $\epsilon$
\begin{equation}
\epsilon = {\epsilon_V   \over 2(d_1 + \epsilon_V)} \left( 3 +  d_1  + \epsilon_V \pm  \left|- 3 + d_1  + \epsilon_V \right|  \right) \, .
\end{equation}
Because there are two roots and an absolute sign two cases need to be examined. Firstly, we assume that 
\begin{equation} \label{eq:sign}
d_1  + \epsilon_V <3 \, ,
\end{equation} 
and this gives the following two solutions 
\begin{equation}  \label{eq:epsilon2}
\epsilon = \epsilon_V\, , \qquad \text{or} \qquad \epsilon = { 3 \epsilon_V   \over d_1  + \epsilon_V } \, .
\end{equation}
Having gradient flow as a possible solution makes sense; recall that the validity of the gradient flow approximation is measured by the smallness of $c_2$ (see e.g.~\cite{Lyth:2009zz,Yang:2012bs}) and whenever it is small we expect the solution to exist. As for the second relation, it yields
\begin{equation}
\epsilon_V < \epsilon,
\end{equation}
which is inconsistent because it would result in negative $\Omega^2$. Therefore, if \eqref{eq:sign} is satisfied then only the gradient flow solution exists. Now we turn into the second possibility, namely
\begin{equation} \label{eq:sign2}
d_1  + \epsilon_V >3 \, ,
\end{equation} 
for which the second solution exists for $d_1>0$ (which comes from the requirement $\epsilon<3$), while the existence of the gradient-flow one requires as usual $\epsilon_V<3$.

\begin{figure}
\centering
\includegraphics[width=0.6\textwidth]{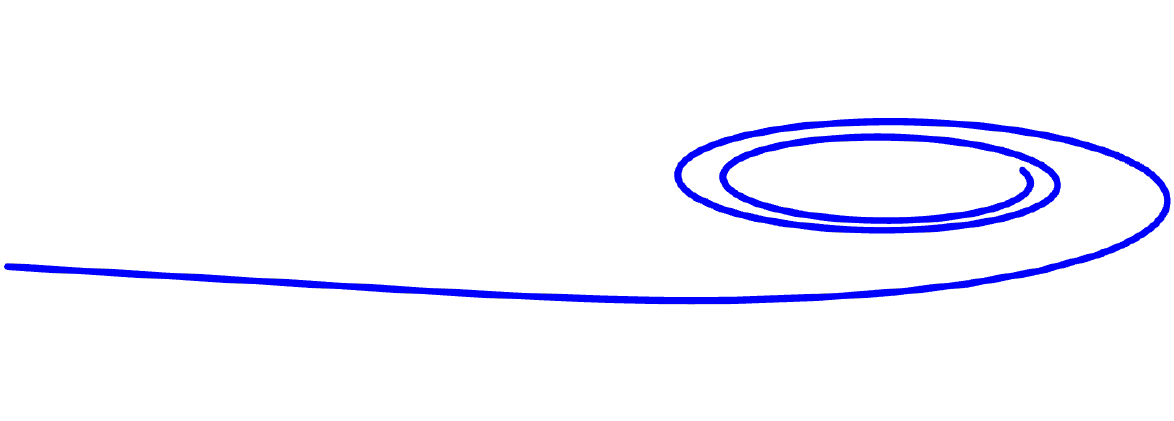}
\caption{An example of a planar trajectory that starts as a line ($\Omega \approx 0$) and ends a spiral with a nonzero slowly-varying turn rate $\Omega$. The torsion is exactly zero in this example. This figure is purely illustrative and does not correspond to any inflationary trajectory we study in this work.}
\label{fig:two_field models}
\end{figure}

Eqs.~\eqref{eq:epsilon1} or \eqref{eq:epsilon2} provide an expression for $\epsilon$ in terms of the two fields $\epsilon = \epsilon(\phi,\chi)$ (for now we take $\phi^k \equiv \left\{ \phi, \chi \right\})$. This, however, does not fix $\epsilon$ completely because if the two fields are treated as independent there will exist a family of solutions for valid initial conditions. If an attractor solution exists the system will choose this specific trajectory which is given in parametric form as e.g.~$\chi(\phi)$. This can be understood using the effective potential description made in \cite{Christodoulidis:2019mkj}: orthogonal fields are almost stabilized at the minimum of their effective potential and only the inflaton remains dynamical. Therefore, to correctly derive the attractor solution we have to find the parametric relation between $\chi$ and $\phi$. As has been noted in \cite{Bjorkmo:2019fls} the parametric relation can be found by constructing different expressions for the turn rate and then match the expressions. Equivalently, we can find a different expression for $\epsilon$ and then compare it with Eqs.~\eqref{eq:epsilon1} or \eqref{eq:epsilon2}. Using $c_1$ and the trace of $w_{\alpha \beta}$ we find the following quadratic equation for $\epsilon$
\begin{equation}
(d_1 + \epsilon_V) \epsilon^2 + {1 \over 2} \left[ c_1 - 2 \epsilon_V (3 + d_1 + \epsilon_V) \right] \epsilon + 3\epsilon_V^2 = 0 \, .
\end{equation}
When $d_1 + \epsilon_V \neq 0$ the two solutions for $\epsilon$ are
\begin{equation} \label{eq:epsilon3}
\epsilon_{\pm} = - {1 \over 2} A  \pm  {1 \over 2} \sqrt{ A^2 - {12 \epsilon_V^2 \over d_1 + \epsilon_V }} \, ,
\end{equation}
where we defined for convenience
\begin{equation}
A=   { c_1 - 6 \epsilon_V   \over 2(d_1 + \epsilon_V)} -  \epsilon_V \, ,
\end{equation}
whereas when $d_1 + \epsilon_V = 0$ \footnote{Note that we have not found any model in the existing literature satisfying this condition.}
\begin{equation}
  \epsilon = {6\epsilon_V^2 \over   6 \epsilon_V - c_1  } \, .
\end{equation}
To better illustrate the generality of our approach, in the next section we will apply the formulae \eqref{eq:epsilon1}, \eqref{eq:epsilon2} and \eqref{eq:epsilon3} on some characteristic models from the literature. Two-field trajectories by definition are given in terms of two basis vectors and so the torsion is identically zero. These curves are planar for Euclidean spaces and in general they lie on the submanifold spanned by the tangent and normal vectors (see Fig.~\ref{fig:two_field models}).

\subsection{Three examples from the literature: hyperinflation, angular and sidetracked inflation} \label{sec:two_field_examples}

Our first example is \textit{hyperinflation} \cite{Brown:2017osf} formulated as follows
\begin{equation} \label{eq:hyperinflation2}
\ud s^2 = \ud \phi^2 + {L^2 \over 4} \sinh^2  \left({\phi \over L}\right) \ud \chi^2  \, , \qquad V = V(\phi) \, .
\end{equation}
This model can be shown to behave very closely to another model with 
\begin{equation} \label{eq:hyperinflation}
\ud s^2 = \ud \phi^2 + e^{2\phi/L} \ud \chi^2  \, , \qquad V = e^{p \phi} \, ,
\end{equation}
for which it is straightforward to check that  
\begin{align}
w_{i} &= (p,0) \, , \\
w_{i;j} &=  e^{2\phi / L } {p \over L} 
\begin{pmatrix}
 0 &0 \\
0 & 1
\end{pmatrix}
\, .
\end{align}
The previous two relations yield $c_1=c_2=0$ so we need to examine when the inequalities \eqref{eq:sign} and \eqref{eq:sign2} are satisfied. For 
\begin{equation}  \label{eq:steepness}
d_1+\epsilon_V={p \over L}  + {1 \over 2} p^2 <3 \, ,
\end{equation}
only gradient flow is possible ($\epsilon = \epsilon_V$ for $p^2<6$), while in the opposite case besides the gradient flow (under the condition $p^2<6$) if additionally $d_1=p /L>0$ holds, we also find the hyperbolic solution \cite{Cicoli:2019ulk,Christodoulidis:2019mkj}
\begin{equation}  \label{eq:epsilon_hyperinflation}
\epsilon = { 3 p L\over 2 + p L} \, .
\end{equation}
For this model $\chi$ is a cyclic variable and hence the solution is expected to be given in terms of $\phi$ only. Therefore, only one expression for $\epsilon$ is needed to fully determine the solution. The hyperbolic solution is a good approximation to the hyperinflation solution in the limit of $\phi\gg L$ and large curvature $L\ll 1$ for which the steepness condition on the potential and $\epsilon$ given in \cite{Brown:2017osf,Mizuno:2017idt,Bjorkmo:2019aev}  as
\begin{equation}
{V_{,\phi} \over V } > 3 L \, , \qquad \epsilon \approx {3 \over 2} {V_{,\phi} \over V} L \, ,
\end{equation}
can be found by expanding Eqs.~\eqref{eq:steepness} and \eqref{eq:epsilon_hyperinflation} for small $L$.

Our second example is \textit{angular inflation} \cite{Christodoulidis:2018qdw}. The model consists of two quadratic fields interacting via a hyperbolic field metric
\begin{equation}
\ud s^2 = {\alpha \over (1 - \chi^2)^2} \left( \chi^2 \ud \phi^2 + \ud \chi^2 \right) \, , \qquad V = {1 \over 2} \alpha \chi^2 f(\phi) =  {1 \over 2} \alpha \chi^2 \left(  \cos^2 \phi + R_{\rm m} \sin^2\phi \right) \, ,
\end{equation}
where $\phi$ and $\chi$ are angular and radial coordinates respectively. For small $\alpha$ it was shown that gradient-flow becomes unsustainable and the system departs to the ``angular'' phase, where the radial coordinate is almost frozen. Defining
\begin{equation}
p \equiv {V_{,\phi} \over V}= {(R_{\rm m} -1 ) \cos \phi \sin \phi \over \cos^2 \phi + R_{\rm m} \sin^2 \phi} \, , \qquad y \equiv {1-\chi^2 \over \alpha} \, ,
\end{equation}
and expanding Eqs.~\eqref{eq:epsilon1} and \eqref{eq:epsilon3} for small $\alpha$ yield 
\begin{equation}
\epsilon \approx {1 \over 2} \alpha p^2 y^2 \, , \qquad  \text{and} \qquad \epsilon \approx {3\alpha y^2 (4 + p^2) \over 6+8y} \, .
\end{equation}
Matching the two expressions provides the parametric relation 
\begin{equation}
1 - \chi^2 \approx {3 \alpha \over p^2} = {3(\cot\chi+R_{\rm m}\tan \chi)^2 \over 4(R_{\rm m}-1)^2}\, ,
\end{equation}
that was found in \cite{Christodoulidis:2018qdw} for this specific model.

Finally, we consider \textit{sidetracked inflation} \cite{Garcia-Saenz:2018ifx}. We assume the ``minimal geometry" and a sum-separable potential
\begin{equation}
\ud s^2 = \left( 1 + 2 {\chi^2 \over L^2 } \right) \ud \phi^2 + \ud \chi^2 \, , \qquad V = U(\phi) +  {1 \over 2} M^2 \chi^2 \, .
\end{equation}
During the evolution, the gradient flow phase becomes geometrically destabilized for sufficiently small enough $L$ and a new phase called ``sidetracked'' emerges which is characterized by an almost constant $\chi$. For this model defining 
\begin{equation}
p \equiv {U_{,\phi} \over V}\, , 
\end{equation}
and assuming $L \ll1 $ then using Eq.~\eqref{eq:epsilon1} we find
\begin{equation}
\epsilon \approx {1 \over 2}{p^2  \over 1+ 2{ \chi^2 \over  L^2} } \, .
\end{equation}
In order to consistently neglect higher order terms we need to further assume $V_{,\chi} \ll V$.  Now using Eq.~\eqref{eq:epsilon3} and expanding around $L=0$ we find an alternative expression for $\epsilon$
\begin{equation}
\epsilon \approx {3 \over 4} {V_{,\chi} L^2 \over V \chi} \left(1+ 2{\chi^2 \over L^2} \right) \, .
\end{equation}
For our choice of a quadratic potential for the heavy field, assuming ${1 \over 2} M^2 \chi^2\ll U$ and equating the two expressions for $\epsilon$ yields the parametric relation
\begin{equation}
1 +  2 {\chi^2 \over L^2} \approx \sqrt{{2 \over 3}}{|p| \sqrt{ U} \over LM} \, ,
\end{equation}
that was found in \cite{Garcia-Saenz:2018ifx}.

\subsection{Two-field metrics with isometries}\label{sec:2f_isometries}
In the case of a metric with isometries there is a natural basis consisting of the Killing vector and its orthogonal vector 
$\{\boldsymbol{K}, \boldsymbol{M} \}$, from which we can construct the orthonormal basis denoted by $\{\boldsymbol{k},\boldsymbol{m}\}$. In the Killing basis the metric is independent of the isometric field and additionally the off-diagonal components of the metric can be set to zero through the diffeomorphism invariance of the metric. Therefore, a generic 2D metric with an isometry can be described by the following simple line element
\begin{equation} \label{eq:isometry}
\ud s^2 = f^2(\chi) \ud \phi^2 + \ud \chi^2 \, ,
\end{equation}
where we identified $\phi$ as the isometric field, and the equations of motion for the two fields are
\begin{align}
&\phi''+ (3- \epsilon)(\phi'+w^{\phi}) + 2{f_{,\chi} \over f} \chi' \phi'=0 \, , \\
&\chi''+ (3- \epsilon)(\chi'+w^{\chi}) -f f_{,\chi} (\chi')^2=0 \, .
\end{align}
In this basis, the unit Killing vector has components
\begin{equation}
k^i \equiv {K^i \over \lVert K^i\rVert} =\left({1 \over f},0 \right) \, ,
\end{equation}
and along with the orthogonal vector $e_{\chi}^i =(0,1)$  (which coincides with the basis vector in the $\chi$ direction) the velocity vector $v^i \equiv \ud \phi^i/\ud N$ is decomposed as
\begin{equation}
v^i = u k^i + v_{\chi} e_{\chi}^i \, ,
\end{equation}
where $u \equiv k_i v^i$ and $v_{\chi} = v^{\chi}$. The components of the velocity are related to the slow-roll parameter as $2\epsilon = u^2 + v_{\chi}^2 $ and they satisfy the following system of equations
\begin{align}
&u'+ (3- \epsilon)(u+w_k) + {f_{,\chi} \over f} u v_{\chi} =0 \, , \\
&v_{\chi}'+ (3- \epsilon)(v_{\chi} + w_{\chi} )  -{f_{,\chi} \over f} u^2=0 \, ,
\end{align}
with the additional definition $w_{k} \equiv w^i k_i$. A slow-roll-like solution consistent with $u' \approx v_{\chi}' \approx 0$ has one of the following forms:
\begin{enumerate}
\item $v_{\chi} \approx - w_{\chi}$ and $u \approx -w_k$. For generic field metric with $f_{,\chi}\neq 0$ this solution is possible only if $w_k \approx 0$, while for the Euclidean space with $f=1$ the solution exists for generic $w_k$. In both cases the resulting solution has small turn rate.

\item $u \approx -w_k$ and $v_{\chi} \approx 0$, so the orthogonal field is frozen. This is possible when $ (3-\epsilon) w_{\chi} \approx (\ln f)_{,\chi} u^2$ is satisfied. The resulting solution is gradient flow if $w_{\chi} \approx (\ln f)_{,\chi}  \approx 0$, otherwise it belongs to the rapid-turn regime.

\item Both velocities can be non-zero for either Euclidean or hyperbolic space (because in that case $f_{,\chi}/f$ is constant). The Euclidean space has only slow-turn solutions, while the hyperbolic one can support solutions with large turn rate, which can be found by solving the equations of motion for $v_{\chi}$ and $u$. The latter equations can be solved consistently for $w_{\chi}=0$ \footnote{To show this one can look for solutions with both fields evolving, $\chi',\phi'\neq0$, $w_{\chi}=c_1$ and $w_k=(\ln V)_{,\phi}/f=c_2$ with $c_1,c_2$ some constants. The latter gives the form of the potential as $V\propto \text{exp}(c_2 f \phi)$ which also yields $w_{\chi}=(\ln V)_{,\chi}=f_{,\chi}$ and this can not be constant unless the space is flat. Therefore, for the hyperbolic space the gradient along $\phi$ has to be zero in order for a rapid-turn solution to exist.}  resulting to the hyperinflation scenario \eqref{eq:hyperinflation} where the hyperbolic solution does not proceed along the isometric ($\phi$) or non-isometric ($\chi$) fields. However, as was shown in \cite{Christodoulidis:2019mkj}, in  a coordinate system different than the original global Poincar\'e coordinates the hyperbolic solution can be shown to be explicitly aligned with another isometry direction. Even though the rapid-turn solution does not proceed along the manifest isometry direction when the parameterization \eqref{eq:hyperinflation} is considered, with a different parameterization of the hyperbolic space inflation does proceed along another isometry. This becomes possible because the hyperbolic space has three Killing vectors which are associated with the three different possible functions $f$ in Eq.~\eqref{eq:isometry} for which the Ricci scalar is constant and negative. Each time the Killing vector can be chosen as the appropriate basis vectors and the symmetry will be explicit in the metric. Therefore, even in the hyperbolic space rapid-turn solutions also proceed along the isometry direction (see also the discussion in App.~\ref{app:hyper} on how to reach this conclusion).

\end{enumerate}
Our previous simple classification scheme proves that whenever an isometry is present then rapid-turn solutions can be realized only along the isometry direction(s), which was an implicit assumption so far in the literature.

\subsection{Metrics without isometries and the stability conditions} \label{subsec:2D_stability_conditions} 

Following the case with isometries it is natural to ask whether rapid-turn solutions can exist for generic geometries without isometries. To investigate this question we first move to a coordinate system where the orthogonal field is frozen and write the most general metric in two dimensions (after some possible field redefinitions) as
\begin{equation}
\ud s^2 =f^2\ud \phi^2+ g^2 \ud \chi^2\, ,
\end{equation} 
where $f,g$ depend on both fields. The equations of motion for the normalized velocities $y\equiv f v^{\phi}$ and $x\equiv g v^{\chi}$ are
\begin{align}
&\phi' = {y \over f} \, , \\ \label{eq:dy}
&y' = - (3-\epsilon) \left(y+{w_{\phi} \over f}\right) - {f_{,\chi} \over g f} x y + {g_{,\phi} \over g f} x^2   \, , \\
&\chi' = {x \over g} \, , \\
&x' = - (3-\epsilon) \left(x+{w_{\chi} \over g}\right)  - {g_{,\phi} \over g f} x y + {f_{,\chi} \over g f} y^2 \, ,
\end{align}
with $w_{\phi}\equiv (\ln V)_{,\phi}$ and $w_{\chi}\equiv (\ln V)_{,\chi}$. In terms of these variables the slow-roll parameter becomes
\begin{equation}
\epsilon = {1\over 2} \left( y^2 + x^2 \right) \, .
\end{equation}Note that when the metric has an isometry the previous normalized velocities become the normalized projections $u$ and $u_{\chi}$ that we used in the previous section. We assume that the frozen (attractor) solution has the form $x=0,y=$ const and $\chi=\chi_0$, where $\chi_0$ is given as a solution to 
\begin{equation} \label{eq:frozen_condition}
(3-\epsilon)w_{\chi}= (\ln f)_{,\chi}2\epsilon \, .
\end{equation}
The latter has exactly the same form as the isometric case, so the absence of the isometry does not affect the existence of such solutions. The slow-roll parameter $\epsilon$ becomes $2\epsilon=w_{\phi}^2/f^2$, which implies that sustained attractor solutions require $(w_{\phi}/f)'\approx 0$. If we let quantities depend on $\phi$ then the attractor solution given as a `critical point' of this dynamical system makes sense only if this dependence is weak. For example, if the metric has an isometry in the $\phi$ direction and the potential has a product-separable form with an exponential in $\phi$ then the $y,\chi,x$ subspace becomes independent from $\phi$ and the stability of the solution can be inferred in the usual way. In fact, by a change of variables one can show that the space decouples even further. 

Defining a new variable $z\equiv y+w_{\phi}/f$ its equation of motion becomes
\begin{equation} \label{eq:zequation}
z' = - (3-\epsilon) z -  {f_{,\chi} \over g f} x z + {g_{,\phi} \over g f} x^2  + {w_{\phi,\chi} \over f}  {x \over g} + {1 \over f} \left( {w_{\phi} \over f} \right)_{,\phi} \left( z - {w_{\phi} \over f} \right)  \, .
\end{equation}
In Eq.~\eqref{eq:dy} the solution $y=-w_{\phi}/f$ requires only $x=0$, leaving $\chi$ unspecified. Therefore, for $z=0$ to be a solution of \eqref{eq:zequation} when $x=0$ the last term should vanish identically, i.e.~for every $\chi$, and this condition restricts the form of available potentials compatible with attractror behaviour. This is also compatible with what one gets for a slowly-varied solution using the Frenet-Serret equations including $w_{\sigma\sigma}$ and $w_{\sigma n}$ which yield $(w_{\chi}/g)_{,\phi} \approx0$ and $(w_{\phi}/f)_{,\phi}\approx0$ respectively.  By virtue of the second condition, the last term in Eq.~\eqref{eq:zequation} can be discarded when it is small (since even after linearization its derivative will remain small), whereas for the second last term this is not always the case.  Being able to neglect the mixed derivatives of the potential can be fulfilled for product-separable potentials or for sum-separable potentials where inflation is driven by the potential energy of the inflaton (see e.g.~the sidetracked model studied earlier in Sec.~\ref{sec:two_field_examples}). Through the Frenet-Serret equation and the relation \eqref{eq:frozen_condition} this term can be written in different ways using
\begin{equation}
{1\over g} {w_{\phi,\chi} \over f} |_{\chi_0}   \approx - {w_{\chi} \over f g} {g_{,\phi} \over g} |_{\chi_0}  \, , \qquad w_{\chi , \phi} |_{\chi_0}(3-\epsilon) \approx (\ln f)_{,\chi \phi} |_{\chi_0} 2\epsilon\, ,
\end{equation}
thus this term may be important when the metric functions $g$ and $f$ depend on both fields. In what follows we assume that the mixed-derivative term can be neglected which is equivalent to considering $f=f(\chi)$; we will return to this assumption at the end of this section. In this case the equation for $z$ to linear order can be integrated out when $w_{\phi}^2 < 6 f_0^2$, showing that the $z$ variable goes to zero.

The assumption $w_{\phi,\chi}\approx0$ constrains the geometries that can support rapid-turn. Using the definition of the turning vector
\begin{equation}\label{eq:definition_turn_rate}
\Omega^i  = {\uD_N v^i \over  \sqrt{2\epsilon}} - {1 \over 2}\eta t^i \, ,
\end{equation}
the turn rate is found to be
\begin{equation}
\Omega \approx -{3 -\epsilon \over g \sqrt{2\epsilon}} w_{\chi} = - {1\over g } \sqrt{2\epsilon} (\ln f)_{,\chi} \, .
\end{equation}
The previous implies that if $g_{,\phi} \neq 0$ then rapid-turn solutions become unsustainable because the time variation of $\ln\Omega$ is directly related to that of $g$:
\begin{equation}
(\ln \Omega)' \approx - (\ln g)'  \, .
\end{equation}
When $g$ depends on the inflaton only the slow-turn solution is allowed.

Setting $z=0$ in the evolution equations for $x,\chi$ the linearized equations become
\begin{align}
& \delta \chi' = {x \over g} \, , \\
& \delta x' = - g \left( w^{\chi}_{\rm eff}\right)_{,\chi}\big|_{\chi_0} \delta \chi +  \left({w_{\phi}^2 \over 2 f^2} -3 + {w_{\phi} g_{,\phi} \over f g}  \right)  \delta x  \, , 
\end{align}
where
\begin{equation}
w^{\chi}_{\rm eff} \equiv {1 \over g^2}\left[ \left( 3 - {w_{\phi}^2 \over 2 f^2}  \right) w_{\chi} - {w_{\phi}^2 f_{,\chi} \over f^3} \right] \, ,
\end{equation} 
is the \textit{effective gradient} for the orthogonal field (introduced in \cite{Tolley:2009fg,Christodoulidis:2019mkj}), $\chi_0$ is a root of the equation $w^{\chi}_{\rm eff} =0$ (which is assumed to exist in order for the frozen solution to exist) and $\delta \chi\equiv \chi - \chi_0$. Note that the equation for $\phi$ can be discarded by performing the redefinition $N\rightarrow\phi$ and changing $e$-fold derivatives with 
\begin{equation}
{\ud \over \ud N} \rightarrow   - {f^2 \over w_{\phi}} {\ud  \over \ud \phi} \, ,
\end{equation}
for which we can write the system as 
\begin{equation} \label{eq:reduced_system}
 - {f^2 \over w_{\phi}}  X_{,\phi} = - \mathcal{J}_{\rm red}  \cdot X
\, .
\end{equation}
Here we defined the reduced stability matrix around the $\chi=\chi_0,~x=0$ solution 
\begin{equation} 
X \equiv \begin{pmatrix}
\delta \chi \\ \delta x
\end{pmatrix} \, , \qquad \mathcal{J}_{\rm red} \equiv  \begin{pmatrix}
0 & -{1 \over g} \\
g \mathcal{M}_{\rm eff} &3 - \epsilon + B
\end{pmatrix}\, ,
\end{equation}
with $\mathcal{M}_{\rm eff}$ denoting the linearization of the effective gradient and $B$ related to the norm of the basis vector  $e_{\chi}^i = \partial_{\chi}$
\begin{equation} \label{eq:M_B}
\mathcal{M}_{\rm eff} \equiv w^{\chi}_{\rm eff , \chi} ={1 \over g^2} {V_{,\chi\chi} \over H^2} + 3 \Omega^2 - {2 f_{,\chi\chi} \over g^2 f} \epsilon \, , \qquad B \equiv  - {w_{\phi} g_{,\phi} \over f g}   = {g_{,\phi} \over g} v =  \left(\ln\norm{ e_{\chi}^i}\right)' \,  .
\end{equation}
Since our `time' variable is $\phi$ we consider the behaviour of the system for $\phi>\phi_0$ if $w_{\phi}<0$ and $\phi<\phi_0$ for $w_{\phi}>0$. As we explained, for a rapid-turn solution with $w_{\phi,\chi} \approx 0$ the metric functions $f$ and $g$ should be independent of $\phi$ (except for the special case of the hyperbolic solution \eqref{eq:hyperinflation}) and $\mathcal{J}_{\rm red}$ has constant elements with $B=0$. The stability will be inferred by simply calculating the eigenvalues of $\mathcal{J}_{\rm red}$; these have non-negative real part when $\mathcal{M}_{\rm eff} >0$, where we also find $\mathcal{M}_{\rm eff}  = (\mu_{\rm eff}/H)^2$ (i.e.~the effective mass of isocurvature perturbations on superhorizon scales). The stability for slow-turn solutions with $g=g(\phi)$ is more tricky because the reduced matrix becomes $\phi$-dependent. Diagonalizing the stability matrix as $\mathcal{J}_{\rm red}=U^{-1}\cdot D_{\mathcal{J}} \cdot U$ the system \eqref{eq:reduced_system} can be written in terms of $Y\equiv U\cdot X$
\begin{equation}
  - {f^2 \over w_{\phi}}   Y_{,\phi} =  D_{\mathcal{J}} \cdot Y  - {f^2 \over w_{\phi}}   U_{,\phi} \cdot U^{-1} \cdot Y \, .
\end{equation}
Assuming that the last term is bounded, i.e.~$\int \ud \phi |U_{,\tilde{\phi}} \cdot U^{-1} |<\infty$ (see e.g.~\cite{nonauton} for more details), the linearized perturbations evolve according to $Y\propto \text{exp} \left( \int \ud \phi (-w_{\phi}/f^2)D_{\mathcal{J}} \right)$. Therefore, for stable slow-turn solutions it is sufficient to demand the eigenvalues of $\mathcal{J}_{\rm red}$ to be non-negative, recovering the two conditions mentioned in \cite{Christodoulidis:2019mkj,Christodoulidis:2019jsx} 
\begin{equation} \label{eq:stab_criteria}
\mathcal{M}_{\rm eff} >0 \, ,\qquad 3 - \epsilon + B>0 \, .
\end{equation}
It is worth mentioning that these criteria do not necessarily imply that the effective mass of superhorizon orthogonal perturbations should be positive. This can be understood using the simple example of the hyperbolic space with $g\propto e^{\phi/L}$ and a symmetric potential. For the slow-turn solution $\mathcal{M}_{\rm eff}$ vanishes identically and hence the equation for $\delta x$ can be directly integrated out, implying that the sign of the second expression in \eqref{eq:stab_criteria} is sufficient to infer stability. Moreover, one finds that $\mu_{\rm eff}^2/H^2 =- B(3 - \epsilon + B)$ \cite{Christodoulidis:2019mkj} and hence if $B>0$ the condition for background stability is fulfilled while the orthogonal perturbation is unstable. This happens for the choice $w_{\phi}/L<0$  as can be checked by using \eqref{eq:M_B}. This simple example shows that the sign of $\mu_{\rm eff}^2$ is not the ultimate stability criterion and its application is justified only in some of the aforementioned cases.

Before concluding the stability section we should return to the key assumption $w_{\phi,\chi}\approx 0$. Although we can not exclude long-lived rapid-turn solutions when the metric functions $f$ and $g$ depend on both fields, these solutions would require the following three relations to hold identically for $\phi$ 
\begin{align}
w_{\phi}(\phi,\chi) \approx  \sqrt{2\epsilon} f(\phi,\chi) \, ,   \qquad w_{\chi , \phi} |_{\chi_0} (3-\epsilon) \approx  (\ln f)_{,\chi \phi} |_{\chi_0}2\epsilon \, , \qquad (\ln f)_{,\chi\phi}|_{\chi_0}  \approx (\ln g)_{,\phi} (\ln f)_{,\chi} |_{\chi_0} \, ,
\end{align}
to ensure that $\chi' , ~\epsilon',~ \Omega' \approx 0$. These solutions, if they exist, they would require a specific interplay between the potential and the field metric, which makes them highly limited. 

Having thoroughly discussed the two-field case we move to three fields.

\section{Three-field solutions} \label{sec:3field}
\subsection{The zero-torsion case} \label{subsec:epsilon}

For three fields the Hessian contains 3 more components: the $_{b \sigma}$ component is related to the torsion, while the $_{nb}$ and $_{bb}$ are not related to kinematic quantities:
\begin{equation}
w_{\alpha \beta} 
\approx
\begin{pmatrix}
{\Omega^2 \over 3- \epsilon} & - \Omega & - {\Omega T \over 3- \epsilon} \\
 - \Omega & w_{n n}  & w_{n b}  \\
 - {\Omega T \over 3- \epsilon} & w_{n b}  & w_{bb} 
\end{pmatrix} \, .
\label{eq:kinematicHessian}
\end{equation}
By analogy to the two-field section, to find an expression for the late-time solution we thus need each time, besides $\epsilon_V$, five more curvature invariants. In contrast to the simple two-field case the resulting $6\times 6$ system of equations can not be solved by standard methods and this forces us to consider some simplifications. One such simplification of geometric origin is to assume zero torsion $T=0$ and with four curvature invariants we are able to successfully solve the system of equations. The solution becomes tractable when accompanied with the slow-roll condition $\epsilon\ll 1 $ and the extreme turning limit $\Omega^2\gg 9$. Using the first four higher order contractions of the Hessian along the gradient vectors we find the following expression for $\epsilon$
\begin{equation}
\epsilon_1 \equiv  {9 \over 2} { c_1^2( c_1^2  -2 c_2 \epsilon_V) \over  c_2^3 + 2 c_1 c_2 (3 c_2-c_3) + c_1^2 (9 c_2 -6 c_3 + c_4) + 2 c_3^2 \epsilon_V - 2c_2 c_4 \epsilon_V}  \, .
\label{eq:invtEps}
\end{equation}
Using the trace of the Hessian $d_1$ instead of $c_4$ we find a second expression
\begin{equation}
\epsilon_2 \equiv  {3 \epsilon_V ( c_1^2  -2 c_2 \epsilon_V) \over c_1^2 (d_1 - 6)  - c_1 c_2 + 2 (c_3 - c_2 (d_1 - 3) ) \epsilon_V}\, ,
\label{eq:invtEps2}
\end{equation}
and finally using additionally $d_2$ instead of $c_3$ we find
\begin{equation}
\epsilon_3  \equiv   {12 \epsilon_V^2 \over -c_1+2 (d_1 - 6)\epsilon_V  - \sqrt{ c_1^2 +4 c_1(d_1 - 6) \epsilon_V - 4\epsilon_V (2 c_2+d_1^2 \epsilon_V - 2 d_2 \epsilon_V) }} \, .
\label{eq:invtEps3}
\end{equation}
These three expressions suffice (at least in theory) to uniquely determine the attractor solution under the previous assumptions. However, in practice we find numerically that even for models that have extreme turn rate the second expression fails to correctly track $\epsilon$. A more accurate expression can be found by solving in terms of the four curvature invariants $c_1,c_2,c_3$ and $d_1$ without assuming $\Omega^2 \gg 9$ or $\epsilon \ll 1$. Solving the system of equations yields two more complicated formulae for $\epsilon$, one of which indeed tracks $\epsilon$ correctly; from now on $\epsilon_{2\rm c}$ will denote that solution. It is worth noticing that one can also find (lengthier) expressions for $\epsilon$ analogous to Eqs.~\eqref{eq:invtEps} and \eqref{eq:invtEps3} without assuming the extreme turning condition.

In the case of $c_1=0$ and $c_2\neq 0$ we can relax the extreme-turning and slow-roll conditions to find the following expression
\begin{equation}
\epsilon = \epsilon_V + {c_2^3 \over 2 c_3^2 - 2 c_2 c_4} \, ,
\end{equation}
assuming that $c_4\neq0$ (because otherwise $\epsilon>\epsilon_V$ which is inconsistent with our earlier requirement $\eta=0$). Using $d_1$ instead of $c_4$ we find the following expression
\begin{equation}
\epsilon = {3 c_2 \epsilon_V  \over c_2(d_1+\epsilon_V) - c_3} \, .
\end{equation}
The latter two relations, along with $c_1=0$, suffice to uniquely determine the parametric relations between the fields.

Lastly, if $c_2=0$ then the column matrix $w_{\alpha \beta}w_{\beta}$ should contain only zero elements (due to positivity of the norm $\norm{ w_{\alpha \beta}w_{\beta}}$: $w_{\alpha} w_{\alpha \beta} w_{\beta \gamma} w_{\gamma} = 0 \Rightarrow w_{\alpha} w_{\alpha \beta}=0$). For the vector $w_{\alpha \beta}w_{\beta} $ we find that its first component is identically zero and, hence, requiring all three components to be zero fixes two components of the Hessian, namely $w_{nn}=3-\epsilon$ and $w_{nb}=T$. Moreover, note that the vanishing of $c_2$ forces every other $c_i$ to be zero and so we can not use them as extra information. In the case of non-zero torsion, calculating $d_n$ for $n>1$ we find that they become linearly dependent and can not provide a solution for both $\epsilon$ and $T$. Therefore, when $c_2=0$ we can only derive solutions for zero torsion and using $d_1$ and $d_2$ and find the following expressions
\begin{equation} \label{eq:c1c2zero}
\epsilon =  \epsilon_V \, , \qquad \epsilon_{\pm} = {6\epsilon_V \over 2 \epsilon_V + d_1 \pm \sqrt{2 d_2 - d_1^2} }\, .
\end{equation}
The latter two expressions describe inflationary solutions with a few additional restrictions between $d_1,d_2$ and $\epsilon_V$ (which we do not list here) that ensure they satisfy $0<\epsilon_{\pm}<\epsilon_V$ and $\epsilon_{\pm}<3$, in complete accordance to our earlier discussion around Eq.~\eqref{eq:sign2}. 

We will illustrate the validity of all previous formulae with examples from the literature in Sec:~\ref{sec:examples_three_fields}.

\subsection{Non-zero torsion}\label{subsec:non_zero_torsion}
Without assuming zero-torsion, we end up with a complicated algebraic system for the components of the Hessian. However, when calculating the curvature invariants $c_1,c_2,c_3$ we find the particular combination $w_{nb}-T$ in different powers. Therefore, if $w_{nb}\approx T$ then the following curvature invariants become independent of the torsion
\begin{align} \label{eq:c1wnb}
c_1 & = {2\epsilon (w_{nn} - 3 + \epsilon) \Omega^2 \over (3 - \epsilon)^2} \, ,\\ \label{eq:c1wnb2}
c_2 &=  {c_1^2 (3-\epsilon)^2 \over 2\epsilon \Omega^2} \, , \\
c_3 &=  {c_1^2 (3-\epsilon)^3 \left(c_1 (3-\epsilon) + 2 \epsilon \Omega^2 \right) \over 4\epsilon^2 \Omega^4} \, , \\
d_1 &= 3 - \epsilon +w_{bb} +{c_1 (3-\epsilon)^2 \over 2\epsilon \Omega^2} + {\Omega^2 \over 3 - \epsilon} \, .  
\end{align}
Note that $w_{bb}$ is also absent from the first three expressions. Trading the turn rate for $\epsilon$ and $\epsilon_V$ using Eq.~\eqref{eq:eps_epsV} we can use the first two invariants to find an expression for the slow-roll parameter which remarkably becomes identical to the two-field expression of Eq.~\eqref{eq:epsilon1} (or the equivalent one if $c_3$ is used in place of $c_2$), while the torsion is left completely unspecified. This implies that $\epsilon$ will be given by the ``two-field formula'' independently of the magnitude of the torsion.  Note that this does not imply that the problem is two-field because the torsion can be significant (as we will demonstrate explicitly in Sec.~\ref{subsec:n_isometries}). Finally, specifying $d_1$ and $d_2$ the torsion can be given in terms of the previous quantities. 

Alternatively, if $|w_{nb}-T| \ll |-3+\epsilon + w_{nn}|$ then one finds that Eqs.~\eqref{eq:c1wnb}-\eqref{eq:c1wnb2} still hold approximately and can therefore be used to derive the same two-field expression for $\epsilon$. As we will show in Sec.~\ref{subsec:n_isometries} all three-field models with 2 isometries share this property (or in general $\mathcal{N}$-field models with $\mathcal{N}-1$ isometry fields).

\begin{figure}
\includegraphics[width=0.7\textwidth]{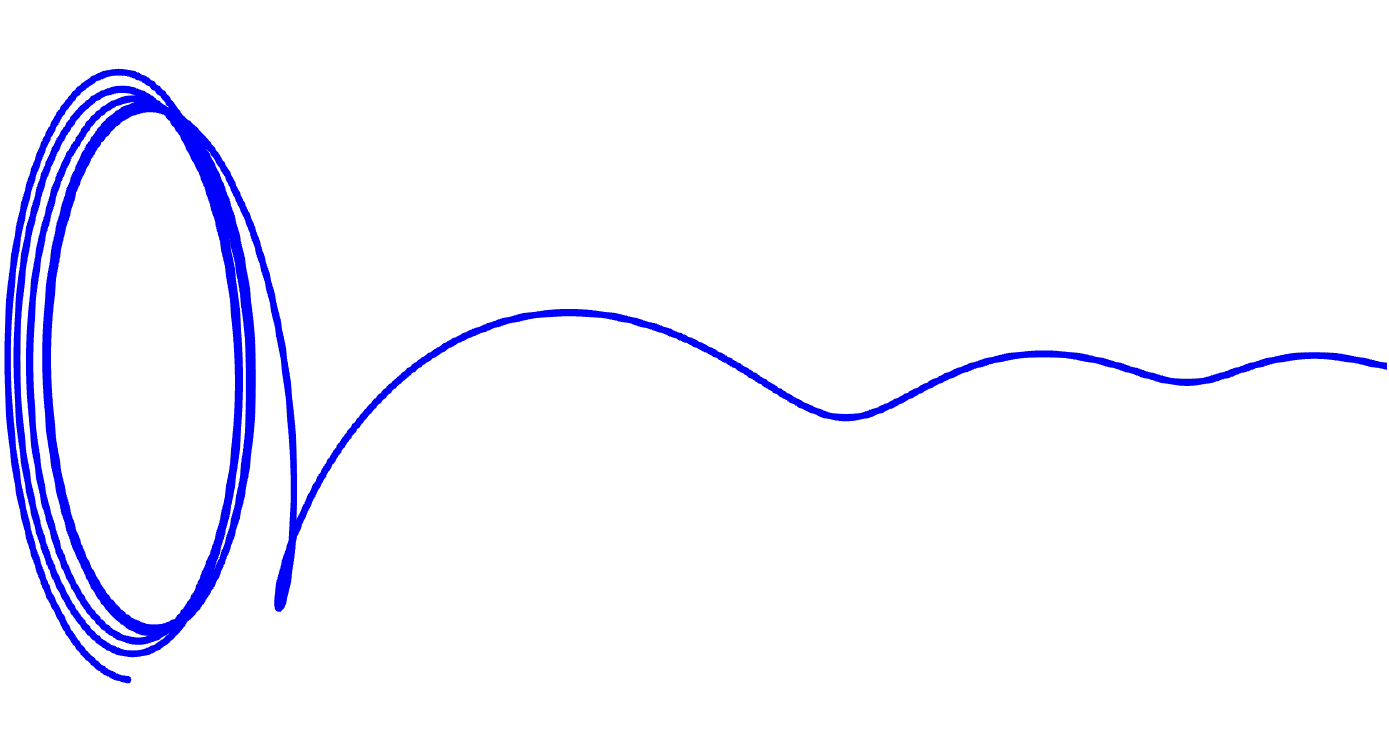}
\caption{An example of a spiraling trajectory with an almost constant turn rate and increasingly larger torsion. The curve starts with $\Omega \gg T$ and ends in the opposite regime $T\gg\Omega$. Therefore, this trajectory interpolates between the slow-twist and slow-turn limits, even though the torsion is gradually increasing. This figure is purely illustrative and does not correspond to any inflationary trajectory we study in this work.}
\label{fig:three_field_plots_large_torsion}
\end{figure}

Slow-twist models will respect the hierarchy $T \ll \Omega$ or $T = \mathcal{O}(\sqrt{\epsilon},|\eta|)$, whereas for rapid-twist models we expect $T = \mathcal{O}( \Omega)$. The opposite regime, namely $\Omega \ll T$ or $\Omega = \mathcal{O}(\sqrt{\epsilon},|\eta|)$ is excluded as an ill-defined case because we can not properly define the normal vector. For this reason we do not expect to construct any such inflationary models (see Fig.~\ref{fig:three_field_plots_large_torsion}).

\subsection{Examples from the literature} \label{sec:examples_three_fields}

Our first example is the \textit{helix model} \cite{Aragam:2019omo}. The potential and field space metric can be written
\begin{align} \label{eq:helix}
G_{ij} &= \begin{pmatrix}
1 & 0 & -\frac{A}{f} \sin(z/f - \theta) \\
0 & \delta r^2 & \frac{A}{f} \delta r \cos(z/f - \theta) \\
-\frac{A}{f} \sin(z/f - \theta) & \frac{A}{f} \delta r \cos(z/f - \theta) & 1 + \frac{A^2}{f^2}
\end{pmatrix},
&
V = e^{z/R} + \Delta\left(1 - \exp\left[-\frac{\delta r^2}{2\sigma^2}\right] \right).
\end{align}
The potential is an exponential in $z$, save for a helical divot of depth $\Delta$ and width set by $\sigma$. The field space is flat, but is expressed in helix-centered cylindrical coordinates $(\delta r, \theta, z)$, so that $\delta r = 0$ tracks the center of the helix. 

By inspection of the potential we can conclude that the system will flow towards its decreasing values at $( \delta r,z) \rightarrow (0,-\infty)$. Close to the center of the helix, the potential can be approximated by 
\begin{equation}
V \approx e^{z/R} + \Delta \frac{\delta r^2}{2\sigma^2} \, , 
\end{equation}
and assuming the existence of a single-degree solution we can parametrize $\delta r$ as $b e^{z/B}$ where $b$ and $B$ depend on the parameters of the problem.\footnote{Note that the differential equations are not well-defined for $\delta r=0$ because the metric blows up at this point, and so we can parameterize $\delta r$ as a small quantity that converges exponentially as a function of time towards its asymptotic value.} At this point it is not clear which term will dominate for $z\ll0$; looking at the equation of motion for $z$, a solution with $z''\approx 0$ provides the following solution for $z'$:
\begin{equation}
z' \approx - {\sigma^2 e^{z /R} + \Delta a b R e^{z/B}\sin\left( z/ f -\theta \right) \over R \sigma^2(e^{z/R} + \Delta b^2 e^{2z/B})} \, ,
\end{equation}
and we can distinguish between the following cases:
\begin{itemize}
\item If $B<R$ then there will be at least one term which grows as z diverges to minus infinity. This means that if a solution exists this will be kinetic domination.

\item If $B>R$ then only the first term survives and the resulting solution is gradient flow $\epsilon \approx \epsilon_V$.

\item Finally, if $B=R$, we find the scaling solution mentioned in the Appendix A of \cite{Aragam:2019omo}. Note that the solution is not restricted to small $b$ or $f$ and it is a rapid-turn one.
\end{itemize}

This analytically known background solution reduces the trajectory to only one degree of freedom, taking $\delta r = b e^{z/R}$ and $\theta = z/f +c$, with $b,c$ functions of the model parameters. This solution has nonzero and constant torsion (either large or insignificant), turn rate, and $\epsilon$. Depending on the magnitude of $T$ we can use the relevant formulae to derive the expression for $\epsilon$. To lowest order in $b$ and in the $z\rightarrow -\infty$ limit, the analytic solution is matched by (the long version of) \eqref{eq:invtEps}
\begin{align}
\epsilon_1 =  \frac{1}{2R^2} \frac{1}{1+A^2/f^2} + \mathcal{O}(b).
\end{align}
Similarly the non-zero torsion expression, \eqref{eq:epsilon1} matches the known analytic solution in the same limit.

\begin{figure}[h]
\includegraphics[width=0.9\textwidth]{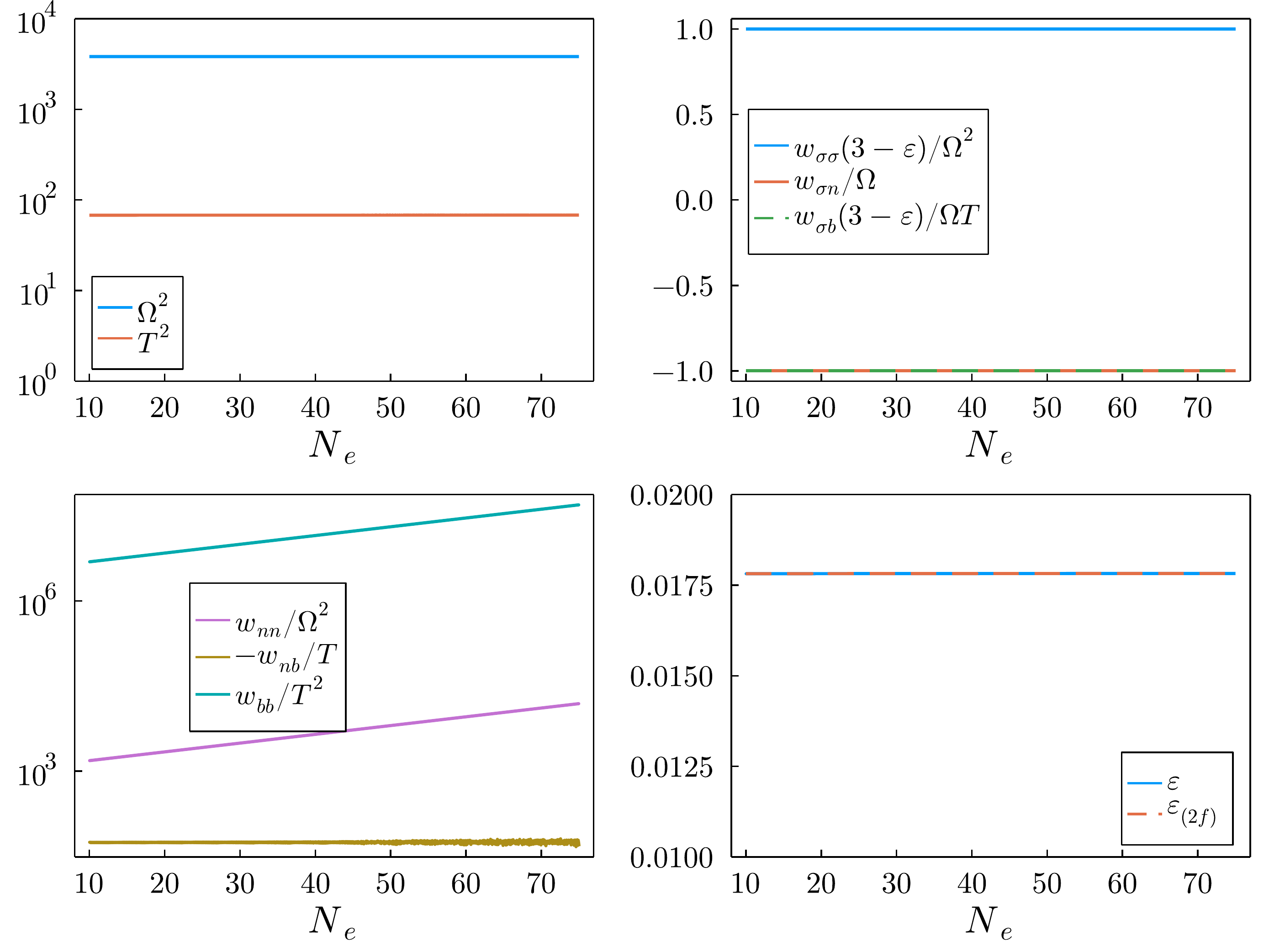}
\caption{(Top left) We show the nearly constant turn rate and torsion on the background solution for the helix model \eqref{eq:helix}. (Top right) We divide the kinematically-known Hessian elements (c.f. \eqref{eq:kinematicHessian}) by their kinematic equivalents, finding excellent agreement. (Bottom left) For completeness, we display the other Hessian elements, with no model-independent kinematic relationships. They are shown divided by kinematic quantities as a convenient parametrization. Note that $|w_{nb}| \gg T$. (Bottom right) We compare the numerical $\epsilon$ to the expressions for $\epsilon$ built out of invariants. Note that despite this being a rapid-turn, rapid-twist model $\Omega \sim \mathcal{O}(10^2), ~T \sim \mathcal{O}(10)$, the two-field expression for $\epsilon$, \eqref{eq:epsilon1}, is in excellent agreement. Parameter values chosen were $\{R,\Delta,A,f,\sigma\} = \{0.7,2.0,3\times10^{-3},4\times10^{-4},10^{-3} \}$, and initial conditions were chosen to be on the steady-state solution, corresponding to $\{\delta r_0,\theta_0,z_0,\delta r_0^\prime,\theta_0^\prime,z_0^\prime \} = \{9.59\times10^{-7},-1249.95, -0.5,-3.42\times10^{-8},-62.38, -0.025 \}$, for an initial and approximately steady-state $\epsilon_0=0.0178$.}
\label{fig:helixEpsilons}
\end{figure}

Another example, is the \textit{$\mathcal{N}$-field hyperbolic problem} \cite{Christodoulidis:2021vye} with
\begin{equation}
\ud s^2 = \ud \phi^2 + \sum e^{k_i \phi} \ud \chi_i^2   \, , \qquad V = e^{p \phi}\, ,
\end{equation}
for which we find $c_1=c_2=0$. Specializing to three fields and applying the formula \eqref{eq:c1c2zero} we find
\begin{equation}
\epsilon= \epsilon_V \, , \qquad \epsilon = {6 p^2 \over 2 p^2 + (k_1+k_2)p \pm |(k_1-k_2) p| } \, ,
\end{equation}
and, hence, we find three possible values for $\epsilon$
\begin{align}
\epsilon= \epsilon_V \qquad &\text{if} \qquad p^2<6 \, , \\
\epsilon = {3 p  \over  k_1 + p } \qquad &\text{if} \qquad  p^2 + k_1 p -6 >0 \, ,~~\text{and}~~ p k_1>0 \, , \\
\epsilon = {3 p  \over  k_2 + p } \qquad &\text{if} \qquad p^2 + k_2 p -6 >0 \, ,~~\text{and}~~ p k_2>0 \, , 
\end{align}
in accordance to the findings of Ref.~\cite{Christodoulidis:2021vye}. Note that the torsion for this model is found to be zero, which as we explained in the previous section is implied by $c_2=0$.

Finally, another specific case that was considered recently in \cite{Aragam:2020uqi} consists of an almost diagonal Hessian in the orthonormal kinetic basis, except for the first $2\times 2$ block, which becomes diagonal in the gradient basis. This is a special case of models described in Sec:~\ref{subsec:non_zero_torsion}, which satisfy $w_{nb}=T=0$ and thus fits our discussion. When the Hessian takes this form, it can be easily checked that the resulting solution for $\epsilon$ is given in terms of quantities of the first block and hence describes a two-field solution. In addition, the first $2\times 2$  block of the perturbations' mass matrix
\begin{align}
    M_{ij} = V_{;ij} - R_{iklj}\dot{\phi}^k\dot{\phi}^l + \left[ (3-\epsilon)\dot{\phi}_i\dot{\phi}_j + \frac{\dot{\phi}_i V_j + V_i \dot{\phi}_j}{H} \right] \, ,
\label{eq:massMatrix}
\end{align}
 decouples and can be studied independently from the others, allowing for a quasi-two-field description.
 Within the remaining $\mathcal{N}-2 \times \mathcal{N} -2$ block, decoupling is not guaranteed: the Hessian structure allows for decoupling but the Riemann tensor term in the mass matrix will spoil it in general, unless the isometries of field space allow for simultaneous diagonalization of the Hessian and the Riemann tensor term.

\section{Specific $\mathcal{N}$-field cases} \label{sec:isometries}
As we explained in Sec.~\ref{subsec:epsilon} finding expressions for the attractor solution (even for  three fields) is a tedious procedure due to the complexity of the multi-field problem. Moreover, following our discussion in the introduction, the attractor solution is characterised by the evolution of one dynamical field, with all other fields frozen at some field value. This behaviour is manifest in a special coordinate system which may be impossible to construct without knowledge of the solution. Thus, pursuing the $\mathcal{N}$-field case forces us to make certain simplifications. 

To simplify the problem, we will consider spaces with at least one isometry, which will enable us to construct the normalized velocities along the Killing directions. Similarly to the two-field case we will show in the following that rapid-turn inflation can only be realized along (a linear combination of) the Killing directions. 

\textbf{Conventions for Latin indices:} early capital letters ($A,B,C,\cdots$) refer to components associated with the isometry directions and early lower case letters ($a,b,c,\cdots$) are reserved for non-isometry fields.

\subsection{Metric with $\mathcal{N}-1$ isometries}  \label{subsec:n_isometries}
We assume $\mathcal{N}$ fields and a metric with $\mathcal{M}$ Killing vectors (with $\mathcal{M} \geq \mathcal{N}-1$).\footnote{Note that the maximum number of Killing vectors for any metric of dimension $\mathcal{N}$ is $(\mathcal{N} + 1)\mathcal{N}/2$.} Moreover, we assume that $\mathcal{N}-1$ of the $\mathcal{M}$ Killing vectors satisfy
\begin{equation}
[\boldsymbol{K}_A,\boldsymbol{K}_B]=0 \, ,
\end{equation}
for $A,B= 1,\cdots,\mathcal{N}-1$ and, hence, we can use them to construct a coordinate basis in which the isometries are manifest. The last basis vector $\boldsymbol{e}_{\chi}$ can be chosen orthogonal to every $\boldsymbol{K}_A$ due to diffeomorphism invariance of the metric (see also App.~\ref{sec:app_cc}). The metric in this basis is independent of every coordinate but the one 
and can be written in block-diagonal form
\begin{equation} \label{eq:isometry_multi}
G_{ij} = 
\begin{pmatrix}
\gamma_{AB}(\chi) & 0_{\mathcal{N}-1\times 1 } \\
 0_{1 \times \mathcal{N}- 1} & 1
\end{pmatrix} \, ,
\end{equation}
generalizing the metric \eqref{eq:isometry}. 
For this geometry we obtain the following non-vanishing Christoffel symbols
\begin{equation}
\Gamma^{\chi}_{\chi j} = \Gamma^{A}_{BC} = 0 \, , \qquad  \Gamma^{\chi}_{BC} = -{1 \over 2} \gamma_{BC,\chi} \, , \qquad  \Gamma^{A}_{\chi B} = - \gamma^{AC}\Gamma^{\chi}_{BC} =  {1 \over 2} \gamma^{AC} \gamma_{CB,\chi}  \, .
\end{equation} 
To find the equations of motion we first note that the Killing equation implies that 
\begin{equation}
v^iv^j (K_{i;j}+K_{j;i}) = 0 \Rightarrow  v^i \uD_N K_{i} = 0 \, ,
\end{equation}
and so the covariant derivative of the Killing vector points in the orthogonal direction $n^i$. Using this result we write the equations of motion for these projections
\begin{equation} \label{eq:killN2}
(K_{Ai}v^i)' = - (3-\epsilon) K_{Ai}(v^i + w^i) \, .
\end{equation}
In terms of the normalized Killing vectors $k_A^i \equiv K_A^i/ \lVert K_A^i \rVert$, where $\lVert K_A^i \rVert = \sqrt{\gamma_{AA}}$, we obtain
\begin{equation} \label{eq:killN3}
u_A' + \left( \ln \lVert K_A^i \rVert \right)_{,\chi} u_A v_{\chi}+ (3-\epsilon )(u_A + \tilde{w}_A)  = 0 \, ,
\end{equation}
with $\tilde{w}_A \equiv k^i_Aw_i$, and the equation for the orthogonal projection 
\begin{equation}
v_{\chi}'  - {1 \over 2} \gamma_{AB,\chi} v^A v^B  +  (3-\epsilon )(v_{\chi} + w_{\chi}) = 0 \,.
\end{equation}

\begin{enumerate}

\item First, we look at the possibility of inflation along the non-isometric field. This requires vanishing gradients along the isometry directions $w_A \approx0$ and the solution has negligible turn rate.

\item Next, we turn our attention to the case of $v_{\chi} \approx 0$. In order for this to satisfy the equation of motion for the orthogonal projection the relation
\begin{equation} \label{eq:consistency_isometry}
{1 \over 2} \gamma_{AB,\chi} v^A v^B  \approx   (3-\epsilon ) w_{\chi}  \, ,
\end{equation}
should be satisfied. For this type of solutions the velocity and acceleration vectors are
\begin{equation}
v^i \approx  \left( - w^A,0 \right) \, , \qquad \uD_N v^i \approx \left(0,\cdots,\uD_N v^{\chi} \right)\, .
\end{equation} 
Using Eq.~\eqref{eq:definition_turn_rate} the normal vector and the turn rate are found to be
\begin{equation}
\Omega n^i  \approx {\uD_N v^{\chi} \over \sqrt{2\epsilon}}\delta^i_{\chi} \, , \qquad \Omega \approx -{3 -\epsilon \over \sqrt{2\epsilon}} w_{\chi} \, ,
\end{equation}
and, hence, $n^i$ is aligned with the basis vector in the orthogonal direction $\chi$. This further gives
\begin{equation} \label{eq:dnni}
\uD_N n^i  \approx \uD_N \delta^i_{\chi} = \Gamma^{i}_{\chi j}v^j  \, ,
\end{equation}
and the binormal vector is found as
\begin{equation}\label{eq:torsion_multi_n}
T b^i \approx \Gamma^{i}_{\chi j}v^j + \Omega  t^i  =  G^{ij} (w_{j;\chi} -w_{j,\chi}) + \Omega  t^i\, ,
\end{equation}
where in the last we used the equations of motion $v^A \approx -w^A$ and the relation $\Gamma^i_{\chi j}w^j=G^{ik}\Gamma_{k \chi}^lw_l$. Taking the derivative of \eqref{eq:consistency_isometry} in terms of $\phi^A$ we find that the mixed potential derivatives vanish, i.e.~$w_{\chi,A}|_{\chi_0}\approx 0$. Therefore, projecting along the torsion and the rest orthogonal vectors we obtain
\begin{equation}
T \approx w_{nb} \, , \qquad w_{nb_i} \approx 0 \, .
\label{eq:torsion_wnb}
\end{equation}
Interestingly, this equation shows that if the torsion is small then the $_{nb}$ component of the Hessian is also small, and the Hessian admits the block diagonal form similar to the ``aligned Hessian approximation'' of Ref.~\cite{Aragam:2020uqi}. In any case, we find that the late-time rapid-turn solution will proceed along one of the isometry fields and the attractor solution can also be found in a coordinate invariant way given by Eq.~\eqref{eq:epsilon1}.

\item Finally, we investigate solutions with the non-isometric and some of the isometry fields evolving $u_A, v_{\chi} \neq 0$. Eq.~\eqref{eq:killN3} implies that a consistent solution for the evolving isometric requires
\begin{equation}
\gamma_{AA} \propto e^{c_{A} \chi} \, , \qquad w_A = 0\, ,
\end{equation}
for some constant $c_A$; this condition fixes the particular diagonal components of the metric. To find the off-diagonal components we use the equation for the orthogonal projection rewritten as
\begin{equation}
v_{\chi}'  - {1 \over 2} \sum_{I,J} (\ln\gamma_{AB})_{,\chi} \gamma_{AB} v^A v^B  +  (3-\epsilon )(v_{\chi} + w_{\chi}) = 0 \, ,
\end{equation}
which can support solutions with $u_A',v_{\chi}' \approx 0$ only if $w_{\chi}' = 0$ and $\gamma_{AB} v^A v^B \propto h_{AB}u_A u_B$ for some constants $h_{AB}$. Therefore, for generic off-diagonal components the previous implies that only one $v_A$ can be dynamical and the rest zero.  In order to have more dynamical fields it is necessary that the off-diagonal components for every pair of fields satisfy $\gamma_{AB}\propto \sqrt{\gamma_{AA}\gamma_{BB}}$. However, this condition describes diagonal metrics in disguise and it suffices to investigate only diagonal metrics; we conclude that solutions with $u_A,v_{\chi}\neq 0$ are valid only for Euclidean or hyperbolic spaces, where the latter has to be written in the exponential parameterization. Using arguments similar to those presented in Sec.~\ref{sec:isometries} we observe that rapid-turn solutions require $v_{\chi}=0$ for generic geometries and hence inflation proceeds along the isometry directions.

\end{enumerate}

We will now investigate which geometries can support small torsion. In the case of a single dynamical field at late times, e.g.~$v^1$, with every other one almost frozen, we find that the components of the torsion vector are given as
\begin{align}
T b^1 &= \left( \Gamma^1_{\chi 1}\sqrt{2\epsilon} + \Omega \right) t^1 \, , \\ \label{eq:torsion_components}
T b^A &= \Gamma^A_{\chi 1} \sqrt{2\epsilon}t^1  \, , \qquad  A \neq 1 \\ 
T b^{\chi} &= 0  \, .
\end{align}
The first equation implies that either $T=0$ or $b^1=0$. If the metric has certain off-diagonal components non-zero then we conclude that $b^1=0$ and the torsion vector lies in the subspace spanned by the rest fields with components given by Eq.~\eqref{eq:torsion_components}. Alternatively, for diagonal metrics and at least two dynamical isometric fields one finds that the torsion vanishes whenever the corresponding diagonal metric components are equal. Then performing a rotation in the isometry subspace one can map the solution $v^i=(v^1,v^2,\cdots)$ to $\tilde v^i = (\tilde v^1,0,\cdots)$ also inducing some off-diagonal components for the metric whenever the original diagonal metric coefficients are not the same. Therefore, we conclude that slow-twist problems with $\mathcal{N}-1$ isometries require diagonal metrics in the coordinate system where there is only one evolving degree of freedom. Otherwise, the torsion is non-negligible.

\begin{figure}[h!]
\includegraphics[width=0.45\textwidth]{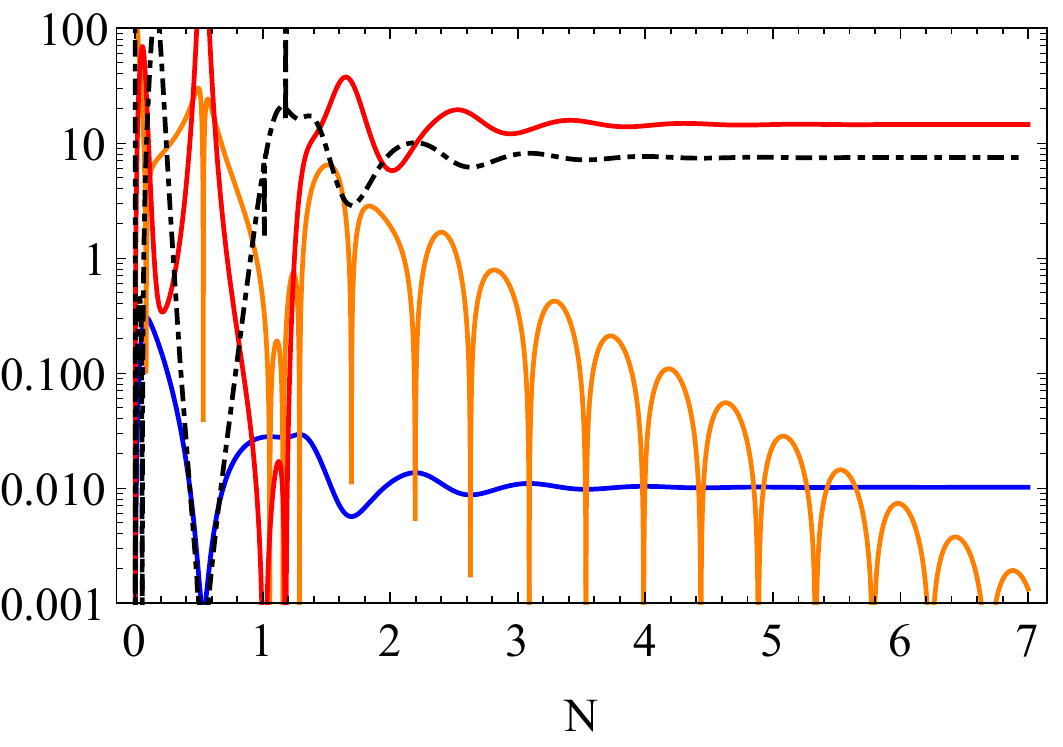}
\includegraphics[width=0.45\textwidth]{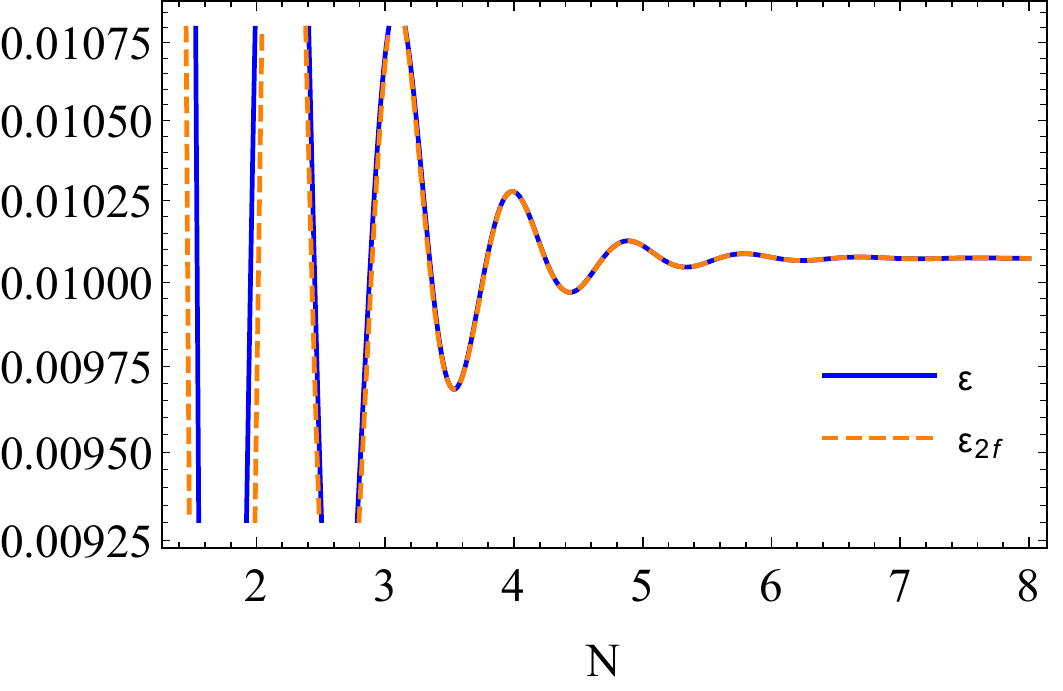}
\caption{A numerical realization with parameters $L=0.01,p_1=1,p_2=6$ and initial conditions $\phi_0=\chi_0=\psi_0=2L$ and $\phi_0'=\chi_0'=\psi_0'=0$. For this model we find significant torsion $T \sim 0.5 \Omega$.
\newline Left: $\epsilon$ (blue), $|\eta|$ (orange), $\Omega^2$ (red) and $T$ (black dotdashed). Right: numerical comparison between $\epsilon$ and formula \eqref{eq:epsilon1} that is valid according to the discussion in Sec:~\ref{subsec:non_zero_torsion}.}
\label{fig:model_3_kinematical}
\end{figure}

As a concrete example, we consider a model with two isometries that satisfies the condition $w_{nb}=T$ 
\begin{equation}
\ud s^2 = \left(1 + { \chi^2 \over L^2}\right) \ud \phi^2 + \cosh \left( {\chi \over L} \right) \ud \psi^2  + \ud \chi^2 \, , \qquad V = \left( 1 + \chi^2 \right) e^{p_1\phi + p_2 \psi}\, .
\end{equation}
For this model an appropriate choice of $p_1,p_2$ and $L$ leads to torsion which is of the same order as the turn rate $T \sim \mathcal{O}(\Omega)$ and therefore to a \textit{rapid-twist} model. In Fig.~\ref{fig:model_3_kinematical} we have plotted various kinematical quantities for the model as well as the different expressions for $\epsilon$. Since the torsion is significant we expect only the formula \eqref{eq:epsilon1} to be valid.

As a final note, we will show that the bending parameter $T_2$ vanishes independently of the magnitude of the torsion. The second binormal vector is obtained from
\begin{equation}
T_2 b_2^i \equiv \uD_N b^i_1 - T n^i \approx {1 \over T} \uD_N \Gamma^{i}_{\chi j}v^j - {1 \over T} \left( \Omega^2 +T^2 \right) n^i = 0 \, ,
\end{equation}
where we made use of
\begin{equation}
 \uD_N \Gamma^{i}_{\chi j}v^j \approx \Gamma^{i}_{m k}v^k \Gamma^{m }_{\chi j}v^j \approx n^i (T^2+\Omega^2) \, .
\end{equation}
This, however, does not provide any further information regarding the other higher order parameters. The vanishing of $T_2$ implies that the kinematics of the problem at the background problem lies on the three-dimensional subspace, spanned by $t^i,n^i$ and $b^i$. In order to complete the orthonormal basis the rest vectors have to be chosen by means of a different method, for instance the Gram–Schmidt process. Nevertheless, in all cases we found that the vanishing of a bending parameter implies the vanishing of every other higher-order parameter as well.

\subsection{Diagonal metric with one isometry} \label{subsec:diagonal_isometry}
We continue the discussion by assuming the existence of an isometry 
\begin{equation}
\ud s^2 = G_{\phi \phi}(\chi^a) \ud \phi^2 + G_{aa}(\chi^a)(\ud \chi^{a})^2\, ,
\end{equation} 
with $\chi$ representing the non-isometry fields. For this geometry there are only three types of non-zero Christoffel symbols
\begin{equation}
 \Gamma^{a}_{\phi\phi} = -{1 \over 2} G^{aa} G_{\phi\phi,a} \, , \qquad  \Gamma^{\phi}_{\phi a}=  {1 \over 2} G^{\phi\phi} G_{\phi\phi, a}  = - G^{\phi \phi} G_{aa}\Gamma^{a}_{\phi\phi}  \, , \qquad \Gamma^a_{aa} = {1 \over 2} G^{aa} G_{aa,a} \, .
\end{equation}
The equations of motion for the normalized projections $u_{\phi} \equiv k_iv^i$ and $u_{a} \equiv \sqrt{G_{aa}}v^a$  and are
\begin{align}
& u_\phi' + {1 \over 2} \left(\ln G_{\phi\phi}\right)_{,a} u_\phi u_a + (3-\epsilon )(u_\phi + w_\phi)  = 0 \, , \\ \label{eq:non_isometry_fields}
& u_a'  - {1 \over 2\sqrt{G_{aa}}} \left(\ln G_{\phi\phi}\right)_{,a}(u_\phi)^2  +  (3-\epsilon )(u_a + w_a) = 0 \, .
\end{align}
In the same spirit as in the previous section we can group solutions according to the following scheme:
\begin{enumerate}
\item Solutions that proceed along at least one of the orthogonal fields while the isometry field is non-dynamical; this is possible only if $w_\phi \approx 0$ and hence the orthogonal fields follow their potential gradient. This describes slow-turn solutions because $\epsilon \approx \epsilon_V$.

\item The inflaton is identified with the isometry direction. This can be either slow- or rapid-turn solution and orthogonal fields are frozen at the minimum of their effective potential, 
\begin{equation}
 \left(\ln G_{\phi\phi}\right)_{,a}  \approx  \sqrt{G_{aa}}(3-\epsilon ) {w_a \over \epsilon} \, ,
\end{equation}
given that all these minima exist.

\item The inflaton, as well as at least one of the orthogonal fields are dynamical. This again requires the potential to be independent of the isometry field, while the metric function should be given as
\begin{equation}
G_{\phi\phi} (\chi^a) \propto e^{c_a \chi^a} \, .
\end{equation}
Note that this metric describes the hyperbolic space and the solution points in the direction of a Killing vector.

\end{enumerate} 
Focusing on the case of inflation along the isometry, the normal vector is given by
\begin{equation}
\Omega n^i =  \left( 0, -  \left(\ln G_{\phi\phi}\right)^{,a} {\sqrt{\epsilon} \over \sqrt{2}}  \right) \, ,
\end{equation}
and we read the normal vector and turn rate as
\begin{equation} \label{eq:omega_epsilon}
n^i \approx \left( 0,-  {\left(\ln G^{\phi\phi}\right)^{,a} \over \sqrt{\left(\ln G_{\phi\phi}\right)^{,b} \left(\ln G_{\phi\phi}\right)_{,b}} }\right) \, , \qquad \Omega^2 \approx {\epsilon \over 2} \left(\ln G_{\phi\phi}\right)^{,a} \left(\ln G_{\phi\phi}\right)_{,a} = (3-\epsilon)^2 {w_a w_a \over 2\epsilon} \, .
\end{equation}
Note that Eq.~\eqref{eq:omega_epsilon} implies that the ratio $\Omega/ \epsilon$ for hyperbolic spaces becomes constant to a great accuracy (see e.g.~\cite{Aragam:2019omo,Christodoulidis:2018qdw} for specific examples where this relation was found numerically). The $i = \phi$ component of the torsion vector is given by
\begin{equation}
T b^{\phi} \approx \Gamma^{\phi}_{\phi j} v n^j +  \Omega t^{\phi} \approx \left(  -\sqrt{{\epsilon \over 2}} \sqrt{\left(\ln G_{\phi\phi}\right)^{,a} \left(\ln G_{\phi\phi}\right)_{,a}}+  \Omega \right) t^{\phi} = 0\, ,
\end{equation}
while for $i\neq \phi$ we obtain
\begin{equation}
T b^{a} \approx \Gamma^{a}_{\phi c} v n^c +  \Omega t^{a}= 0\, ,
\end{equation}
and so we conclude that this type of models belong to the slow-twist class. 

This result (as well as the one from the previous subsection) implies that rapid-twist models require field metrics with off-diagonal elements.

\subsection{$\mathcal{N}$-field background stability} \label{subsec:n_field_stability}

For both classes of models we find that the mixed potential terms vanish, which allows the linearized equation for the isometry field(s) $z_A\equiv u_A+w_A$ to decouple
\begin{equation}
z_A' = - (3-\epsilon) z_A \, ,
\end{equation}
and so $z_A \rightarrow 0$. Moreover, for the rapid-turn solution the norm of the Killing vectors depends only on the frozen field(s) and, hence, it suffices to investigate the matrix of effective gradients. 

\begin{itemize}

\item
For the first class, the $\mathcal{N}-1$ equations for the isometry fields can be ignored and the effective gradient is given by
\begin{equation}
w^{\chi}_{\rm eff}  = \left(3 - {1 \over 2} \gamma^{IJ} w_{I}w_{J} \right) w_{\chi} - {1 \over 2} \gamma_{AB,\chi} \gamma^{AI} \gamma^{BJ} w_{I}w_{J} \, ,
\end{equation}
with effective mass 
\begin{align}
\mathcal{M}_{\rm eff} &= {V_{,\chi\chi} \over H^2}  - w_{\chi} \left[(3-\epsilon) w_{\chi} - {1 \over 2} \gamma_{AB,\chi}  w^A w^B \right] - {1 \over 2} \gamma_{AB,\chi \chi} w^{A} w^{B} + \gamma^{AC} \gamma_{AB,\chi} w^{B}   \gamma_{CD,\chi} w^{D} \, .
\end{align}
With the aid of Eq.~\eqref{eq:dnni} we find that the last term is equal to
\begin{equation}
\gamma^{AC} \gamma_{AB,\chi} w^{B}   \gamma_{CD,\chi} w^{D} = 4  \Gamma^{A}_{\chi B} v^{B} \gamma_{AC} \Gamma^{C}_{\chi D} v^{D} = 4(T^2 + \Omega^2) \, .
\end{equation}
Moreover, the second term is related to the Riemann tensor as
\begin{equation}
R_{\chi AB \chi} w^{A} w^{B} = {1 \over 2} \gamma_{AB , \chi\chi} w^{A} w^{B} + \Gamma^{\chi}_{BC} \Gamma^C_{A\chi} w^{A} w^{B}  \, ,
\end{equation}
and hence the effective mass becomes equal to
\begin{align}
\mathcal{M}_{\rm eff} &= {V_{,\chi\chi} \over H^2}  - R_{\chi AB \chi} w^{A} w^{B} +  3 (T^2+ \Omega^2)= \mathcal{M}_{nn} + 3 (T^2+ \Omega^2) \, .
\end{align}
Note that both the turn rate and the torsion have a stabilizing effect on the effective mass.

\item
In the second class, the linearized equations for the orthogonal fields in the $z$ variables are
\begin{align} 
&(\delta \chi^a) '= u_a \, , \\
& \delta u_a' = w^{a}_{\rm eff,b} \delta u_b-  (3-\epsilon_0)  \delta u_a \, ,
\end{align} 
where the effective gradient vector is given as
\begin{equation}
w^{a}_{\rm eff}  = \left(3 - {w_{\phi}^2 \over 2G_{\phi\phi}} \right) w_a - {1 \over 2} (\ln G_{\phi\phi})_{,a}{w_{\phi}^2 \over G_{\phi\phi}} \, .
\end{equation}
In matrix form the equations are
\begin{equation} \label{eq:effective_matrix_nfields}
\begin{pmatrix}
\delta \chi \\
\delta u
\end{pmatrix} ' = - 
\begin{pmatrix}
0_{\mathcal{N}-1 \times \mathcal{N}-1}  & - I_{\mathcal{N}-1 \times \mathcal{N}-1} \\
(\mathcal{M}_{\rm eff})^a_b  & (3-\epsilon ) I_{\mathcal{N}-1 \times \mathcal{N}-1}
\end{pmatrix}
\begin{pmatrix}
\delta \chi \\
\delta u
\end{pmatrix} \, ,
\end{equation}
The effective mass matrix can be expressed as
\begin{align}
(\mathcal{M}_{\rm eff})^a_b  = &\left[ \epsilon w_{a} (\ln G_{\phi\phi})_{,b} +  \left(3 -\epsilon \right) w_{a,b} - (\ln G_{\phi\phi})_{,ab}\epsilon +  \epsilon (\ln G_{\phi\phi})_{,a}  (\ln G_{\phi\phi})_{,b} \right] \, .
\end{align}
or using the equations of motion 
\begin{align}
(\mathcal{M}_{\rm eff})^a_b  = &\left[ {V_{,ab} \over H^2} - { G_{\phi\phi,ab} \over G_{\phi\phi}} \epsilon +  2\epsilon (\ln G_{\phi\phi})_{,a}  (\ln G_{\phi\phi})_{,b} \right] \, .
\end{align}
Similarly to the three-field case the eigenvalues of the full system can be expressed in terms of the eigenvalues of the effective mass matrix as (see App.~\ref{app:block})
\begin{align}
\lambda_{a,\pm} = {1 \over 2} \left( 3 - \epsilon \pm \sqrt{(3 - \epsilon)^2 - 4 \mu_{a}}\right) \, .
\end{align}
The conditions for stability are
\begin{align}
\text{Im}(\mu_a)^2 < (3 - \epsilon)^2\text{Re}(\mu_a) \, ,
\end{align}
where the eigenvalues of the effective mass are model dependent.
\end{itemize}

\subsection{Remarks on the stability of inflationary backgrounds in supergravity models} \label{subsec:supergravity}

Because string compactifications tend to come with an abundance of light scalar fields, studying string-inspired multi-field inflationary models is a natural target of research.
Much of the relevant literature has focused on supergravity constructions, being a low-energy effective theory description of string theory.
Attempts at constructing single-field supergravity inflation have been historically plagued by the ``$\eta$-problem'', essentially stated as the inability for one scalar field to remain parametrically lighter than the others once quantum corrections are considered \cite{Baumann:2014nda}.
Alternatively, the $\eta$-problem may be viewed as a characteristic failure for single-field supergravity constructions to satisfy the stability criteria discussed earlier.
If we retain the connection to string theory, then parameters in supergravity models are not free to be arbitrarily small or large, and are naturally expected to be $\mathcal{O}(1)$ in Planck units.
This effect contributes to the $\eta$-problem, and in practice makes high-curvature field spaces difficult to construct in multi-field supergravity models \cite{Aragam:2021scu}.
Moreover, there seems to be some confusion in the literature as to the equivalent stability criteria for multi-field supergravity models in the context of inflation, which we 
discuss below. This is because the standard approach utilizes the eigenvalues of the Hessian matrix $K^{\Phi \overline{\Phi}}V_{\Phi \overline{\Phi}}$ to infer the consistency of the single-field truncation or to discover instabilities of the multi-field trajectory  (see e.g.~\cite{Kallosh:2010xz,Achucarro:2012hg}). However, this overlooks contributions from the turn rate and the Riemann curvature tensor.

To understand the origin of this confusion, let us consider the linearized equations of the multi-field problem
\begin{equation}\label{eq:linearized_delta_phi}
\ddot{\delta \phi}^i + 2\Gamma^i_{kj} \dot{\phi^k} \dot{\delta \phi^j}+ 3H \delta \dot\phi^i + 3\delta H \dot\phi^i+ \left(  \Gamma^i_{kl,j} \dot{\phi^k} \dot{\phi^l} + V^{~i}_{,j} \right) \delta \phi^j = 0\, ,
\end{equation}  
from which we observe that besides the Hessian, the `mass' term receives contributions from parts of the Riemann tensor. However, for pure de Sitter solutions defined in the bulk of space, that is for all $\phi^i$, the metric should be finite at these points and terms involving background velocities vanish. This further implies that the de Sitter solutions satisfy
\begin{equation}
\dot{\phi}^i = 0 \, , \qquad V^{,i} =0 \, , 
\end{equation} 
or they correspond to critical points of the potential. Evaluating the linearized equation on these solutions yields
\begin{equation}\label{eq:linearized_critical}
\ddot{\delta \phi}^i + 3H_0 \delta \dot\phi^i +V^{~i}_{,j}\delta \phi^j = 0\, ,
\end{equation}  
or in first order form in terms of the e-folding number
\begin{align}
 (\delta \phi^i)' & = y^i \, , \\
 (y^i)' & = - 3 y^i - 3 {V^{~i}_{,j} \over V} \delta \phi^j \, .
\end{align} 
The stability matrix for this system $Y'=-J \cdot Y$, very similar to Eq.~\eqref{eq:effective_matrix_nfields}, takes the form
\begin{equation}
J = \begin{pmatrix}
0_{\mathcal{N}\times \mathcal{N}} & -I_{\mathcal{N}\times \mathcal{N}} \\
3  \mathcal{H}^{~i}_{j} & 3I_{\mathcal{N}\times \mathcal{N}} 
\end{pmatrix}  \, ,
\end{equation}
where $\mathcal{H}$ is the normalized mixed Hessian
\begin{equation}
\mathcal{H}^{~i}_{j} \equiv {V_{,j}^{~i} \over V} \, .
\end{equation}
Denoting the eigenvalues of $\mathcal{H}^{~i}_j$ as $h_i$ we can use them to find the eigenvalues of the full problem as
\begin{equation}
\lambda_{i,\pm} = {1 \over 2} \left( 3 \pm \sqrt{3 - 4 h_{i}}  \right) \, .
\end{equation}
Note that since $\mathcal{H}$ is symmetric the eigenvalues are real. It is straightforward to check that the condition for stability of the full problem reduces to positivity of the eigenvalues of the Hessian (see App.~\ref{app:block}). Thus, we recover the usual claim regarding de Sitter stability, whenever the solution is defined in the bulk. Moreover, note that on the critical points of the potential the covariant derivatives and normal derivatives coincide. As a side note, we should mention that stable de Sitter solutions can easily be constructed in the simplest supergravity scenarios, e.g.~in $\mathcal{N}=1$ supergravity. The following choice of Kahler potential and superpotential
\begin{equation}
K= \ln \left( 1 + \Phi \overline{\Phi} \right) \, , \qquad  W = \Phi \, ,
\end{equation}
yields a two-field model in the polar representation $\Phi=re^{i\theta}$ with a field space of constant positive curvature and a spherically symmetric potential
\begin{equation}
\ud s^2 = {1 \over (1+r^2)^2} \left( \ud r^2 + r^2 \theta^2 \right)   \, , \qquad V = 1 + 4r^2 + 7r^4 + 4r^6 \, .
\end{equation}
This potential has a global minimum at $r=0$ and so a stable de Sitter solution exists. However, the existence of stable de Sitter minima in string theory remains an open problem (see e.g.~\cite{Andriot:2019wrs}).

On the contrary, for quasi-de Sitter solutions one can not ignore terms proportional to the inflaton's velocity. These terms provide extra contributions to the masses of the orthogonal perturbations (see Sec.~\ref{subsec:2D_stability_conditions}). For inflationary models this has already been appreciated and studied by earlier works, such as \cite{GrootNibbelink:2000vx,GrootNibbelink:2001qt,Peterson:2010np,Peterson:2011ytETAL,Gong:2011uw,Renaux-Petel:2015mga}. In the simplest two-field models with isometries the turn rate, the field-space curvature and the norms of the Killing vectors should also be considered in addition to the Hessian eigenvalues.  Even in the latter case, it could be argued that if the Hessian has negative eigenvalues this could induce an instability to the full system.\footnote{A simple counter-example is the angular inflation model for which the Hessian of the potential has one negative eigenvalue during the angular phase.} To show explicitly why this is not the case we consider the generic form of perturbations around quasi de-Sitter space  
\begin{equation} \label{eq:general_perturbation}
\delta x''+ A \cdot \delta x'+M \cdot \delta x = 0 \, ,
\end{equation}
where $\delta x$ is the column $1\times\mathcal{N}$ and $A,M$ are $\mathcal{N} \times\mathcal{N}$ matrices. The previous equation could describe the linearized equations of background solutions \eqref{eq:linearized_delta_phi} or the equations of motion for the gauge invariant orthogonal perturbations. Moreover, the matrix $A$ is not diagonal, with the  off-diagonal elements related to the non-geodesic motion. Using the same steps as in the App.~\ref{app:block} we find the eigenvalue equation for the system \eqref{eq:general_perturbation} as
\begin{equation}
\text{det}\left(  I \lambda^2 -A  \lambda + M \right) = \text{det}\left(  I \lambda^2 - U\cdot A \cdot U^{-1} \lambda + D_M  \right)  =0 \, ,
\end{equation}
where $U$ is the unitary matrix that diagonalizes $M$. Unless $U$ also diagonalizes $A$ (which is not expected in general), the eigenvalues of the full problem are directly related to the eigenvalues of the mass matrix $M$ only if $A$ is a constant multiple of the identity matrix. Therefore, a negative real part in the eigenvalues of $M$ will not induce instabilities to the full system in general.

Before concluding this subsection we should make the following important distinction: when we discuss the stability of the inflationary trajectory we refer to the path the system follows on its way to the (stable) minimum of the potential. This is not to be confused with the stability of critical point of the potential, which is a point, in contrast to the inflationary trajectory which is a one-parameter curve in field space. A stable critical point $\phi^i_{\rm cr}$ (which is such that no term in the evolution equations diverge) requires positive eigenvalues of the Hessian matrix or the mass term in Eq.~\eqref{eq:linearized_critical}, whereas an attractor inflationary solution requires investigation of a quantity that contains extra terms in addition to the Hessian.

We further comment on specific supergravity models in Sec. \ref{sec:manyfieldssim}.

\section{First order perturbations} \label{sec:observables_stability}

\subsection{Three-field perturbations} \label{subsec:three_field_perturbations}

We will calculate the effect of the torsion on the curvature perturbation. We specialize to $\mathcal{N} = 3$, as higher order turn rates enter the perturbations' equations of motion with higher numbers of fields.
These equations of motion, following \cite{Kaiser:2012ak,Cespedes:2013rda},\footnote{Note that the previous references adopted different sign conventions for $\Omega$ and $T$.} \footnote{See also \cite{Achucarro:2018ngj,Pinol:2020kvw} on how the perturbations' equations generalize to $\mathcal{N}$ fields.} can be written in terms of the $e$-folding number as
\begin{align}
&\mathcal{R}'' + (3 -\epsilon + \eta) \mathcal{R}' + \kappa^2 \mathcal{R}= 2 {\Omega \over \sqrt{2\epsilon}} \left[ Q_n' + \left(3 - \epsilon + {1 \over 2} \eta  + (\ln \Omega)' \right)  Q_n \right]\, ,\\
&Q_n'' + (3 -\epsilon  ) Q_n'  - 2 T Q_b' + \left( \kappa^2+ \mathcal{M}_{nn} - \Omega^2 - T^2\right) Q_n +  \left[- (3 - \epsilon) T - T' + \mathcal{M}_{nb}\right] Q_b= -2 \Omega \sqrt{2 \epsilon} \mathcal{R}'   \, , \\
&Q_b'' + (3 -\epsilon)  Q_b'  + 2 T Q_n' + \left(\kappa^2 + \mathcal{M}_{bb} - T^2 \right) Q_b +  \left[  (3 - \epsilon) T + T' + \mathcal{M}_{nb} \right] Q_n = 0 \, ,
\end{align}
where $\mathcal{R}$ is the curvature perturbation defined as $\mathcal{R}\equiv Q_{\sigma}/\sqrt{2\epsilon}$ and in this section we use $k$ to refer to each mode's wavenumber and $\kappa \equiv k / (aH)$. For the three-field case it is important to first investigate under what conditions the perturbations remain stable. At superhorizon scales, using the time derivative of the curvature perturbation \cite{Bassett:2005xm,Malik:2008im}
\begin{equation}
\mathcal{R}' = - {1 \over \epsilon} \kappa^2 \Psi +\sqrt{{ 2 \over  \epsilon}} \Omega Q_n \, ,
\end{equation}
where $\Psi$ is the Bardeen potential 
 one finds that $Q_n$ decouples from $\mathcal{R}$ and, hence, orthogonal perturbations can be studied separately. Ignoring prime derivatives of background quantities, isocurvature perturbations obey
\begin{equation}
\begin{pmatrix}
Q_n \\
Q_b
\end{pmatrix}
'' +
\begin{pmatrix}
3 - \epsilon & -2T \\
2T &3 - \epsilon
\end{pmatrix}
\begin{pmatrix}
Q_n \\
Q_b
\end{pmatrix}
' + 
\begin{pmatrix}
 \mathcal{M}_{nn} + 3 \Omega^2 - T^2 & -(3 - \epsilon)T  + \mathcal{M}_{nb} \\
 (3 - \epsilon) T + \mathcal{M}_{nb} &   \mathcal{M}_{bb} - T^2
\end{pmatrix}
\begin{pmatrix}
Q_n \\
Q_b
\end{pmatrix}
\approx 0 \, .
\end{equation}
In order for these perturbations to decay on superhorizon scales, namely for the solution $Q_n,Q_b,Q_n',Q_b' \rightarrow 0$, the eigenvalues of the $4\times 4$ system should have non-positive real part. Applying the Routh-Hurwitz criterion and defining the following matrix 
\begin{equation} \label{eq:m_sh}
\mathcal{M}_{\rm sh} \equiv
\begin{pmatrix}
\mathcal{M}_{nn} + 3 \Omega^2 - T^2&   - (3 - \epsilon)T + \mathcal{M}_{nb} \\
  (3 - \epsilon) T + \mathcal{M}_{nb}  & \mathcal{M}_{bb}  - T^2
\end{pmatrix} \, ,
\end{equation}
we obtain the following inequalities in terms of the trace and determinant of $\mathcal{M}_{\rm sh}$
\begin{equation} 
\begin{aligned}
 & 4T^2 +  \text{tr}(\mathcal{M}_{\rm sh}) > 0 \, , \\
& 0 < 4 \text{det}(\mathcal{M}_{\rm sh}) < \left( 4T^2 +  \text{tr}(\mathcal{M}_{\rm  sh}) \right) \left( 4T^2 +  \text{tr}(\mathcal{M}_{\rm sh}) + 2 (3 - \epsilon)^2 \right)  \, .
\end{aligned}
\label{eq:superhorizon_criteria}
\end{equation}
Although these conditions are not very enlightening when $T\neq0$, we observe that the effect of a large torsion is mostly stabilizing whenever it also dominates over other quantities.

On sub-Hubble scales, the torsion and the masses interact to allow for the perturbations' growth in various scenarios. We defer to the analyses of \cite{Aragam:2023adu,perseas} in this case.

\subsection{Two case examples}

\subsubsection{Model 1}\label{subsubsec:model2}
\begin{figure}[h!]
\centering
\includegraphics[width=0.9\textwidth]{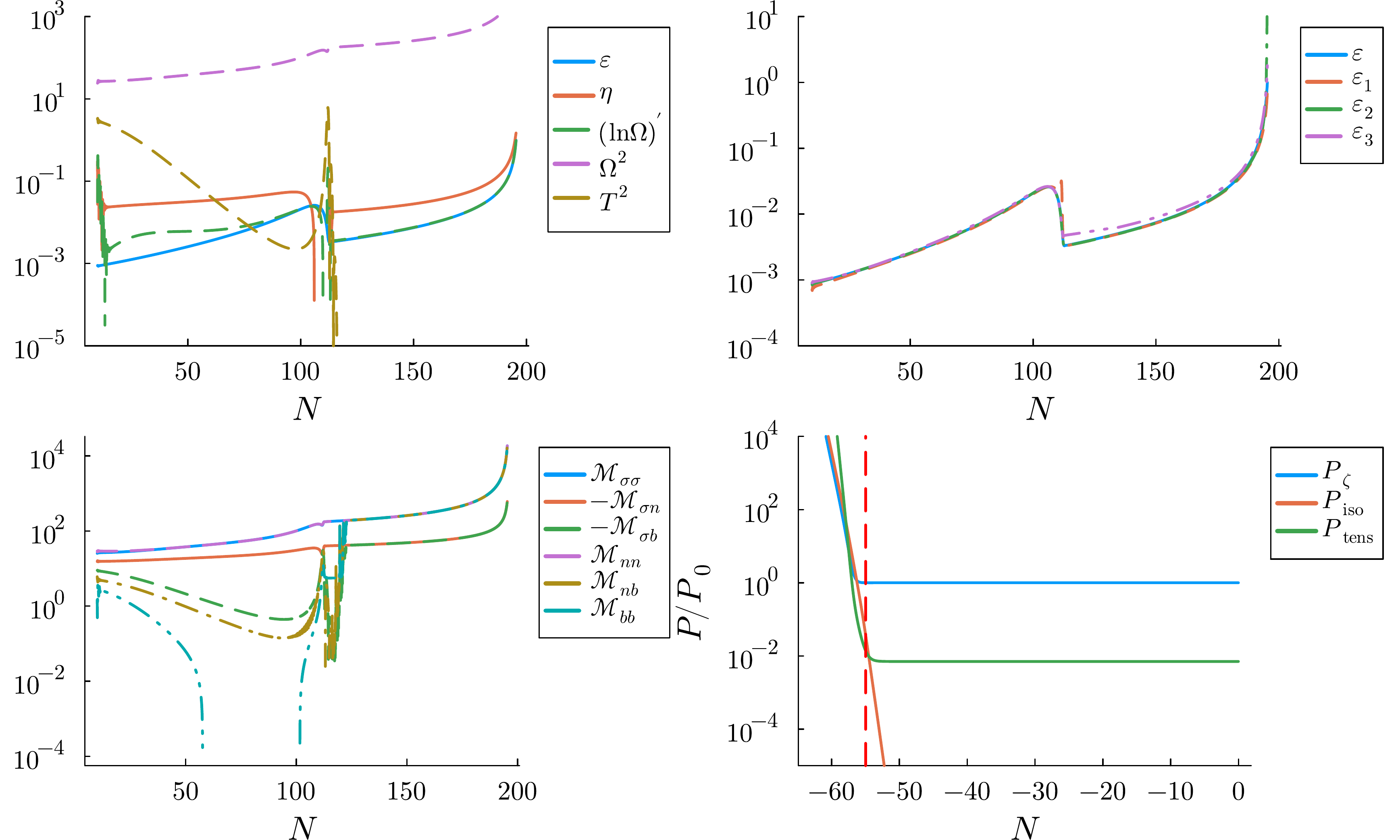}
\caption{A numerical realization with parameters $L_1=0.005, L_2=0.001,M=6$ and initial conditions $\phi_0=4,\chi_0=2,\psi_0=4$ and $\phi_0'=\chi_0'=\psi_0'=0$. We omit the initial 10 e-folds of evolution approaching the shown attractor for the clarity of the figure.
In the top left panel, we plot the background quantities $\epsilon,\eta,\Omega^2,T^2$ and $(\ln \Omega)'$. The model undergoes a slow-roll, rapid turn trajectory for $\sim 195$ e-folds. The model has a rapidly decaying torsion until the transient at $N\sim110$ e-folds, at which point the torsion decays to zero within numerical precision.  In the top right panel, we display a numerical comparison of various expressions for $\epsilon$ of Sec:~\ref{subsec:epsilon}. All agree well during the entire evolution.
In the bottom left panel, we display the components of the mass matrix during the background evolution. Before the transient, we can see that $\mathcal{M}_{\sigma\sigma}\sim \Omega^2 + \mathcal{O}(\epsilon,\eta)$ and $\mathcal{M}_{\sigma n}\sim -3\Omega + \mathcal{O}(\epsilon,\eta)$ as expected from \eqref{eq:kinematicHessian}. Similarly, we observe that $\mathcal{M}_{\sigma b} \sim -\Omega T$. We also notice $\mathcal{M}_{nb}\sim T$, which we expect from \eqref{eq:torsion_wnb}. After the transient, the dynamics become quasi-two-field, and $(\mathcal{M}_{\sigma b},\mathcal{M}_{nb},\mathcal{M}_{bb})\rightarrow (\mathcal{M}_{\sigma n},\mathcal{M}_{\sigma\sigma},\mathcal{M}_{\sigma\sigma})$. In the bottom right panel, we can see the adiabatic, isocurvature, and tensor powerspectra at the pivot scale of $k_\star = 0.002$ Mpc$^{-1}$ evolving as a function of $N$. We imposed Bunch-Davies initial conditions 8 e-folds before horizon exit, finding that more sub-horizon evolution did not affect $n_s$. The isocurvature powerspectra decay to the limits of numerical precision rapidly after horizon exit (vertical red line). This model's observables match Starobinksy inflation well, with its predictions lying in the center of Planck's $n_s$ and $r$ plane: $n_s = 0.968, r=7.05\times 10^{-3}$ at the end of inflation. Note that, although we only plot $P(k_\star,N)$, to measure $n_s$ we compute the scalar perturbations at a range of scales around the pivot scale, see Appendix \ref{sec:numerics} for details on our numerical procedure. }
\label{fig:model_2_all}
\end{figure}

We consider a generalization of the sidetracked model with a double Starobinsky potential and `minimal geometry'. The field metric and the potential are given as follows
\begin{equation}
\ud s^2 = \left(1 + { \chi^2 \over L_1^2}\right) \ud \phi^2 +\left(1 + { \chi^2 \over L_2^2} \right) \ud \psi^2 +  \ud \chi^2  \, , \qquad V = {1 \over 2}M^2 {\chi}^2 + \left( 1 - e^{-{\sqrt{2/3}\psi}} \right)^2 + 2\left( 1 - e^{-{\sqrt{2/3}\phi}} \right)^2 \, .
\end{equation}
In Fig.~\ref{fig:model_2_all} we have plotted the basic kinematical quantities for the model and the different expressions for $\epsilon$, as well as the mass matrix and the predicted power spectra. The plot illustrates the behaviour of the torsion according to the discussion in Sec.~\ref{subsec:n_isometries}. Initially, both $\phi'$ and $\psi'$ are non-zero and contribute to a non-zero, albeit decreasing, torsion. However, $\phi'$ decreases faster and reaches zero at around $110$ $e$-folds at which point the torsion drops to zero as well. This also causes $\epsilon$ to drop as it now receives contribution solely from the gradient of $\psi$.

\subsubsection{Model 2} \label{subsubsec:model1}
We move to an alternative generalization of the sidetracked model with
\begin{equation}
\ud s^2 = \left(1 + { \chi^2 \over L_1^2}\right)\left(1 + { \psi^2 \over L_2^2}\right)  \ud \phi^2 + \ud \chi^2 + \ud \psi^2 \, , \qquad V = {1 \over 2}m_{\chi}^2 {\chi}^2 + {1 \over 2}m_{\psi}^2 {\psi}^2 + \left( 1 - e^{-{\sqrt{2/3}\phi}} \right)^2 \, .
\end{equation}

During the sidetracked phase, assuming $L_1 \ll L_2$, the kinetic basis is approximately aligned with the field basis: $\hat{t}^i \sim -\hat{\phi}^i, \hat{n}^i \sim -\hat{\chi}^i, \hat{b}^i \sim \hat{\psi}$.

This allows us to calculate the torsion in terms of the background trajectory, using \eqref{eq:torsionNormFS}. We find

\begin{align}
\Omega &\simeq \frac{m_\chi^2 \chi}{\sqrt{2\epsilon}} \, , \\
T = \frac{1}{\Omega}\frac{V_{b\sigma}}{H^2}&\simeq  \frac{\sqrt{2\epsilon}}{m_\chi^2 \chi} \frac{2\alpha \psi (e^{-\alpha\phi}-e^{-2\alpha\phi})}{L_1 L_2 + L_1 \psi^2 + L_2 \chi^2 + \chi^2 \psi^2} \sqrt{ \frac{L_1 L_2 (L_1 + \chi^2)}{L_2+\psi^2} } \, ,
\end{align}
where $\alpha=\sqrt{2/3}$. A local maximum in $\psi$ exists when $\psi = \sqrt{L_2/2}$, giving $T = \frac{1}{\Omega} \frac{4\alpha (e^{-\alpha\phi} + e^{-2\alpha\phi}) L_1}{3\sqrt{3} \sqrt{L_1 L_2 (L_1+\chi^2)}}$, which is heavily suppressed as long as $L_1, L_2$ are small and $\chi$ is large.

\begin{figure}[t]
\includegraphics[width=\textwidth]{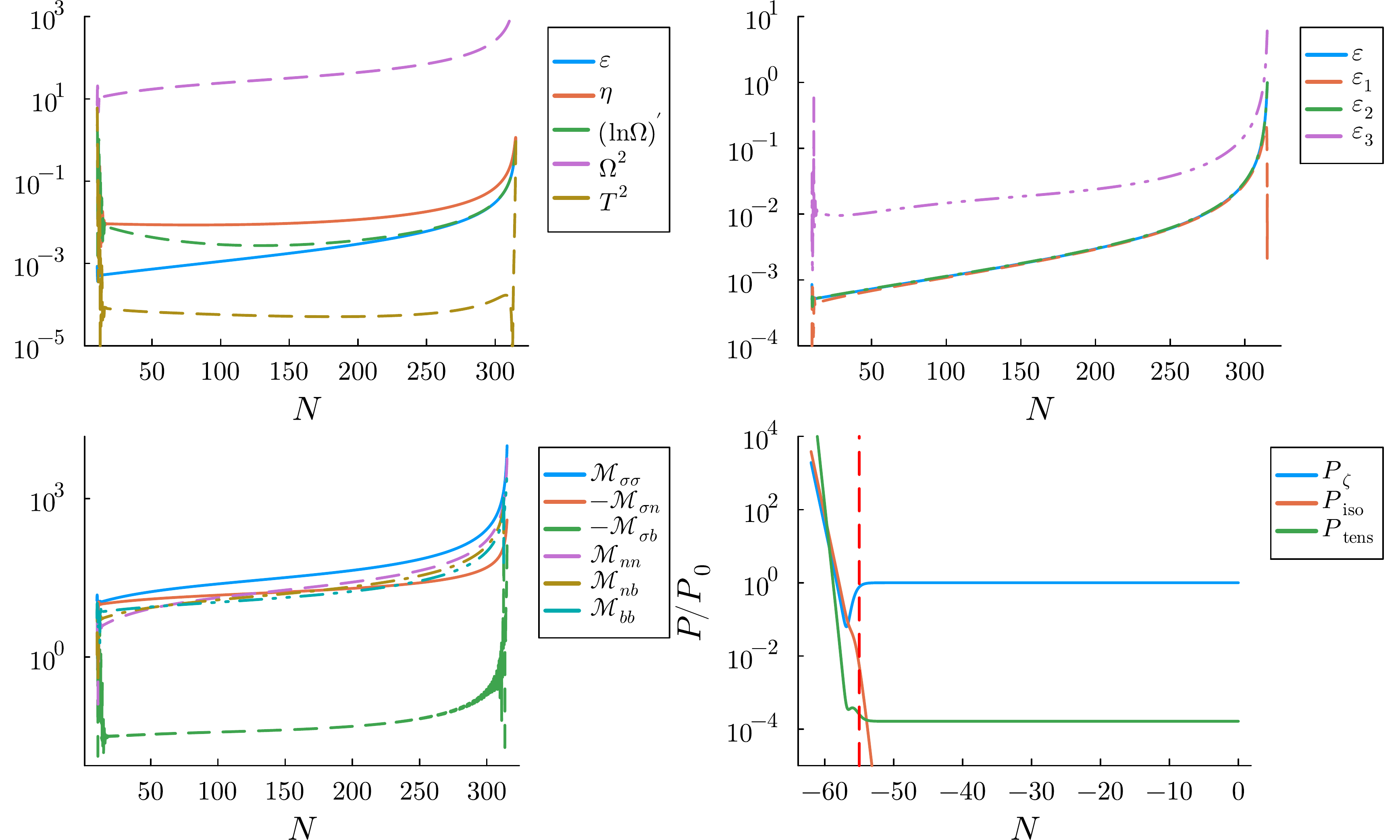}
\caption{A numerical realization with parameters $L_1=0.005, L_2=0.01, m_{\chi}=3,m_{\psi}=1$ and initial conditions $\phi_0=4,\chi_0=2,\psi_0=4$ and $\phi_0'=\chi_0'=\psi_0'=0$.
\newline Top Left: $\epsilon,\eta,\Omega^2,(\ln \Omega)'$, and $T^2$. Top Right: numerical comparison of various expressions for $\epsilon$ of Sec:~\ref{subsec:epsilon}. Bottom Left: Comparison of the mass matrix elements. Bottom Right: The scalar and tensor power spectra at the pivot scale as a function of $N$. The brief period of subhorizon growth biases the prediction of $n_s$ away from the single-field Starobinsky value, reaching $n_s=1.005$. The isocurvture power spectra remain negligible, decaying to numerical precision shortly after horizon exit. Note that, although we only plot $P(k_\star,N)$, to measure $n_s$ we compute the scalar perturbations at a range of scales around the pivot scale, see Appendix \ref{sec:numerics} for details on our numerical procedure.}
\label{fig:model_1_all}
\end{figure}

In Fig.~\ref{fig:model_1_all} we have plotted various kinematical quantities for the model, different expressions for $\epsilon$, the mass matrix elements, and the perturbations' powerspectra.
We note that the observed torsion is very small on the attractor, with $T \lesssim 10^{-2}$, matching our analysis.
The agreement between $\epsilon$ and $\epsilon_3$, the approximation of \eqref{eq:invtEps3}, can increase significantly by including another complicated expression that is derived without the extreme turning condition.

The powerspectra experience a brief transient growth before horizon exit, making this model's observables substantially different than the predictions from single-field Starobinsky inflation. We can understand this growth by studying the perturbations in the WKB approximation, following \cite{Aragam:2023adu}. Because the torsion is so small, we can neglect the third field's effects on the perturbations, and recover the two-field result that sub-horizon growth will occur when $k/k_\star < \Omega \sqrt{1-\xi_{nn}} $, where $\xi_{nn} \equiv \mathcal{M}_{nn}/\Omega^2$, assumed to be constant in the WKB approximation \cite{Fumagalli:2020nvq}. Then, in the $T=0$ limit, we expect sub-horizon growth whenever $\mathcal{M}_{nn}<\Omega^2$ around the time of horizon exit. For the case we present in Figure \ref{fig:model_1_all}, $\mathcal{M}_{nn}/\Omega^2 \sim 0.6$ at the time of horizon exit, predicting around two e-folds of sub-horizon growth at the rate of $\mathcal{R} \propto e^{i q N}$, with $q^2 = (k/a H)^2 + \Omega^2 (3+\xi_{nn})/2 - \Omega \sqrt{(2 k/a H)^2+(3+\xi_{nn})^2 \Omega^2 / 4}$.
The average growth exponent for the plotted parameters is $x \sim 1.4$ predicting around an order of magnitude of $P_\zeta \propto e^{2x}$ growth, which matches the figure.

\FloatBarrier
\subsection{Many-field simulations}
\label{sec:manyfieldssim}
To explore a model with many fields and a high number of isometries, we chose the following metric and potential
\begin{align}
G_{ij} = \mathrm{diag}\left(1,e^{\lambda_1 \chi}, e^{\lambda_2 \chi}, e^{\lambda_3 \chi}, \ldots, e^{\lambda_{\mathcal{N}-1} \chi} \right) \, , \qquad V = 1 + \sum_{i=1}^{\mathcal{N}} \left( A_i \cos{(k_i \phi^i)} + B_i \sin{(k_i \phi^i)} \right) \, , 
\label{eq:manyfieldssim}
\end{align}
where the $\lambda_i$, $A_i$, $B_i$, and $k_i$ are all free parameters of the model. The metric is entirely a function of one field, but with different functions in each coordinate -- we therefore have $\mathcal{N}-1$ metric isometries by construction.

Finding long-lived rapid-turn inflationary initial conditions for this model is difficult, given a total of $6\mathcal{N}-1$ free parameters in $\vec{\phi}(t=0)$, $\dot{\vec{\phi}}(t=0)$, $\vec{\lambda}$, $\vec{A}$, $\vec{B}$, and $\vec{k}$. We detail our full search strategy of this parameter space in Appendix \ref{sec:blackboxing}, and successfully found several points of interest in parameter space.

\begin{figure}[h!]
\centering
\includegraphics[width=0.9\textwidth]{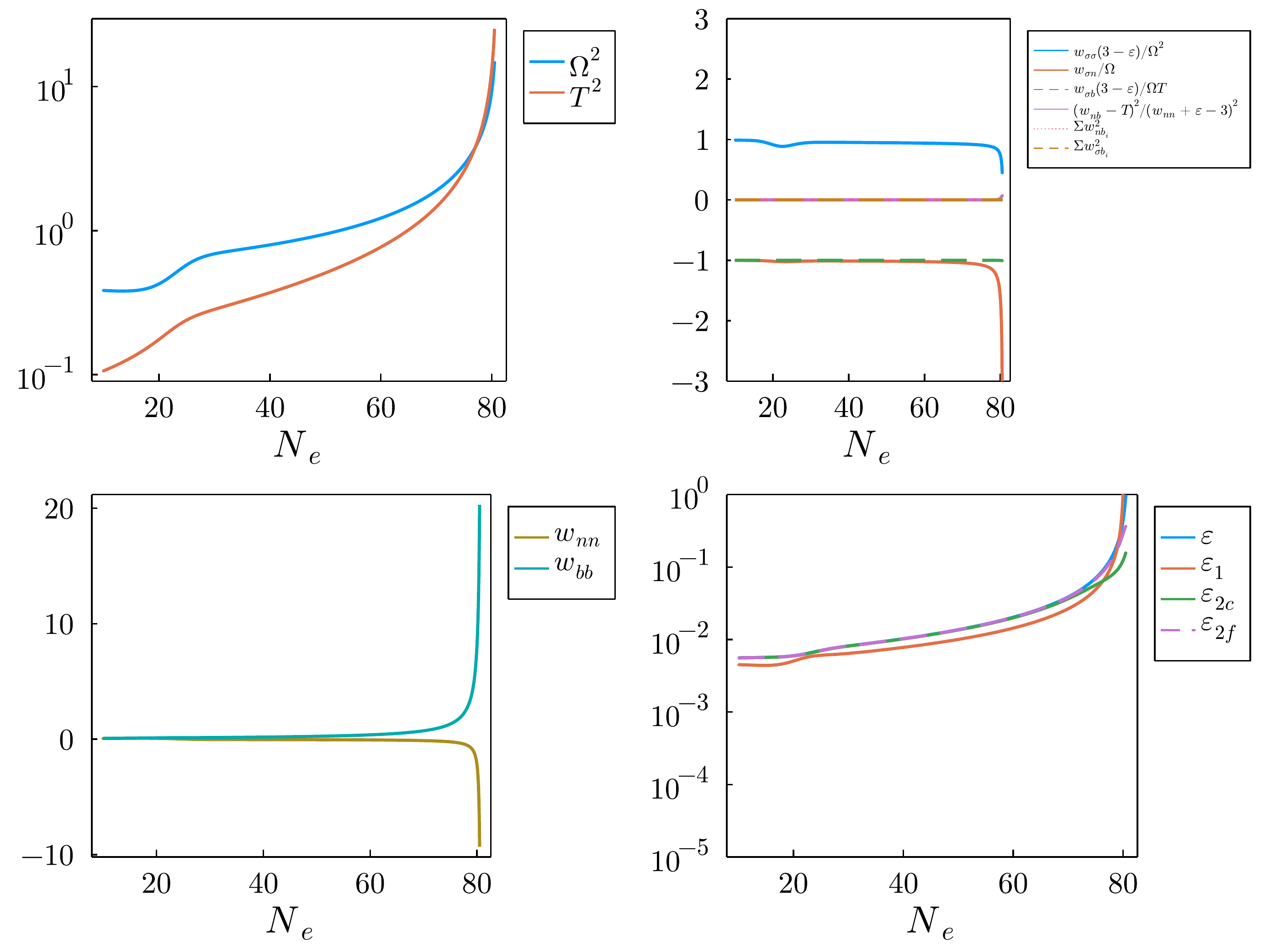}
    \caption{We study a slow-roll, moderate-turn trajectory for the model \eqref{eq:manyfieldssim}. Despite $\mathcal{O}(1)$ turn rate, many of our rapid-turn predictions for $\epsilon$ and the Hessian elements agree well. The parameter values chosen are $\mathcal{N}=4$, $\dot{\phi}^i(t=0)=0$, $\{ \vec{\phi},\vec{\lambda},\vec{A},\vec{B},\vec{k} \} \simeq \{$ $0.311$, $0.947$, $5.057$, $8.753$, $19.237$, $14.513$, $1.807$, $0.111$, $6.610$, $19.151$, $0.212$, $12.849$, $12.300$, $8.118$, $6.798$, $4.933$, $19.430$, $5.200$, $3.443\}$. (Top left) The turn rate and torsion for the model. (Top right) Elements of the Hessian divided by their kinematic relationships, when known. The fourth and higher basis vectors are labeled $b_i$. (Bottom left) Some of the other Hessian elements with no known kinematic relationships. We do not display any of the unknown Hessian elements past the first $3\times 3$ block. (Bottom right) We compared the numerical $\epsilon$, in blue, to several of the invariant expressions presented in this paper. The two-field expression (dotted, purple) overlaps the numerical $\epsilon$ (in blue).}
\label{fig:manyfieldssim}
\end{figure}

We plot the evolution of one of them in Fig \ref{fig:manyfieldssim}, with $\mathcal{N}=10$. Even though the evolution is 10-dimensional field space, this model is slow-twist so many of the expressions for $\epsilon$ agree well with the numerics, and match our expectations from Sec \ref{subsec:n_isometries}.

Another model inspired by supergravity and constructed in \cite{Aragam:2021scu} has the K\"{a}hler and superpotential
\begin{align}
K &= -3\alpha \log\left(\phi + \bar{\phi} \right) + S \bar{S} \\
W &= S(p_0 + p_1 \phi)
\end{align}
where $\phi,S$ are complex scalar fields. The equivalent field space metric and potential for the real fields $\phi = r_1 + i r_2$, $S= r_3 + i r_4$ are
\begin{equation}
\begin{aligned}
V = \frac{8^{-\alpha } {r_1}^{-3 \alpha } e^{{r_3}^2+{r_4}^2}}{3\alpha} &\times \Bigg[
3 \alpha  {p_0}^2+3 \alpha  \Bigg(
{p_0}^2 \left({r_3}^2+{r_4}^2-1\right) \left({r_3}^2+{r_4}^2\right) + 2 {p_0} {p_1} {r_1} \left({r_3}^4+{r_3}^2 \left(2 {r_4}^2-3\right)+{r_4}^4-3 {r_4}^2+1\right) \\
&+ {p_1}^2 \left({r_1}^2 \left({r_3}^4+{r_3}^2 \left(2 {r_4}^2-5\right)+{r_4}^4-5 {r_4}^2+1\right)+{r_2}^2 \left({r_3}^2+{r_4}^2-1\right) \left({r_3}^2+{r_4}^2\right)+{r_2}^2\right) \Bigg)\\
&+9 \alpha ^2 \left({r_3}^2+{r_4}^2\right) \left(({p_0}+{p_1} {r_1})^2+{p_1}^2 {r_2}^2\right)+4 {p_1}^2 {r_1}^2 \left({r_3}^2+{r_4}^2\right)\Bigg]
\end{aligned}
\label{eq:poly1alphapotential}
\end{equation}
and 
\begin{align}
g_{ij} = \text{diag}\left( \frac{3\alpha}{2r_1^2}, \frac{3\alpha}{2r_1^2},2,2  \right),
\label{eq:poly1alphametric}
\end{align}
where we allow for a nonzero $S$ field.

This form of field space is quite similar to \eqref{eq:isometry_multi}, with $r_1$ as the isometry field.
We therefore expect that any late-time rapid-turn attractors will necessarily contain a component of the velocity in the $r_1, r_2$ subspace. Additionally, we expect no off-diagonal terms of the metric in the Killing basis so we expect this model to be slow-twist.

\begin{figure}[h]
\centering
\includegraphics[width=\textwidth]{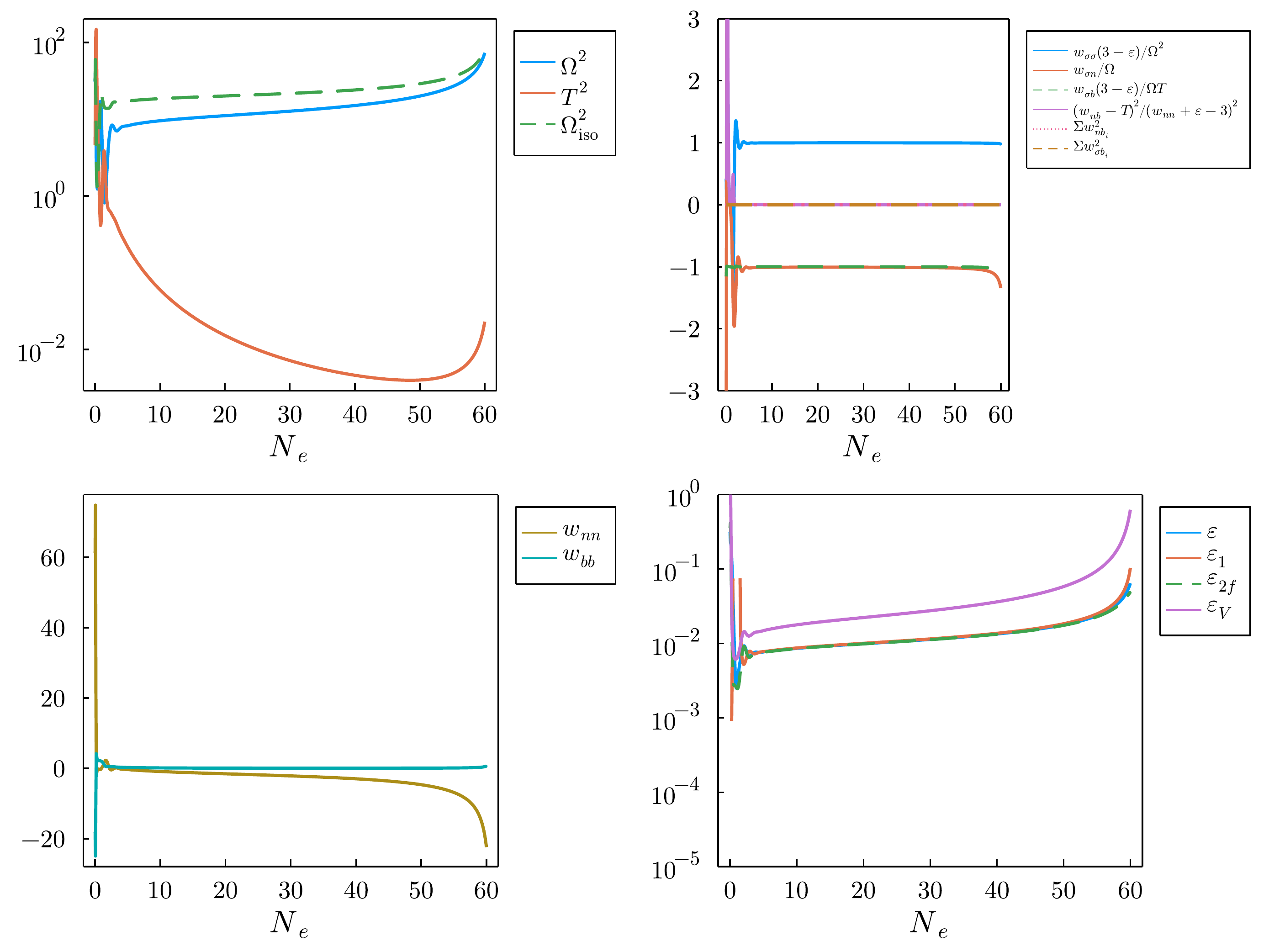}
\caption{A rapid-turn solution in the model \eqref{eq:poly1alphapotential}, \eqref{eq:poly1alphametric}. While initially large, torsion rapidly decays. We compare the numerical turn rate to the one expected from the isometry structure of the model, $\Omega_\mathrm{iso}$, c.f. \eqref{eq:omega_epsilon}. The field values over time can be seen in figure \ref{fig:poly1alphafields}. The initial conditions chosen were $\{\vec{\phi}$, $\vec{\phi}^\prime$, $\alpha$,$p_0$,$p_1\} = \{0.167$, $7.945$, $0.0250$, $0.001753$, $-0.493$, $-1.004$, $-0.267$, $-0.0439$, $0.000206$, $99.69$,$-12.407\}$.}
\label{fig:poly1alpha}
\end{figure}

After an extensive scan for rapid-turn initial conditions, we choose a point in parameter space and plot this model's background evolution in figure \ref{fig:poly1alpha}. We confirm the majority of the field motion to be in the $(r_1,r_2)$ subspace and the torsion to be small.

\begin{figure}[h]
\centering
\includegraphics[width=\textwidth]{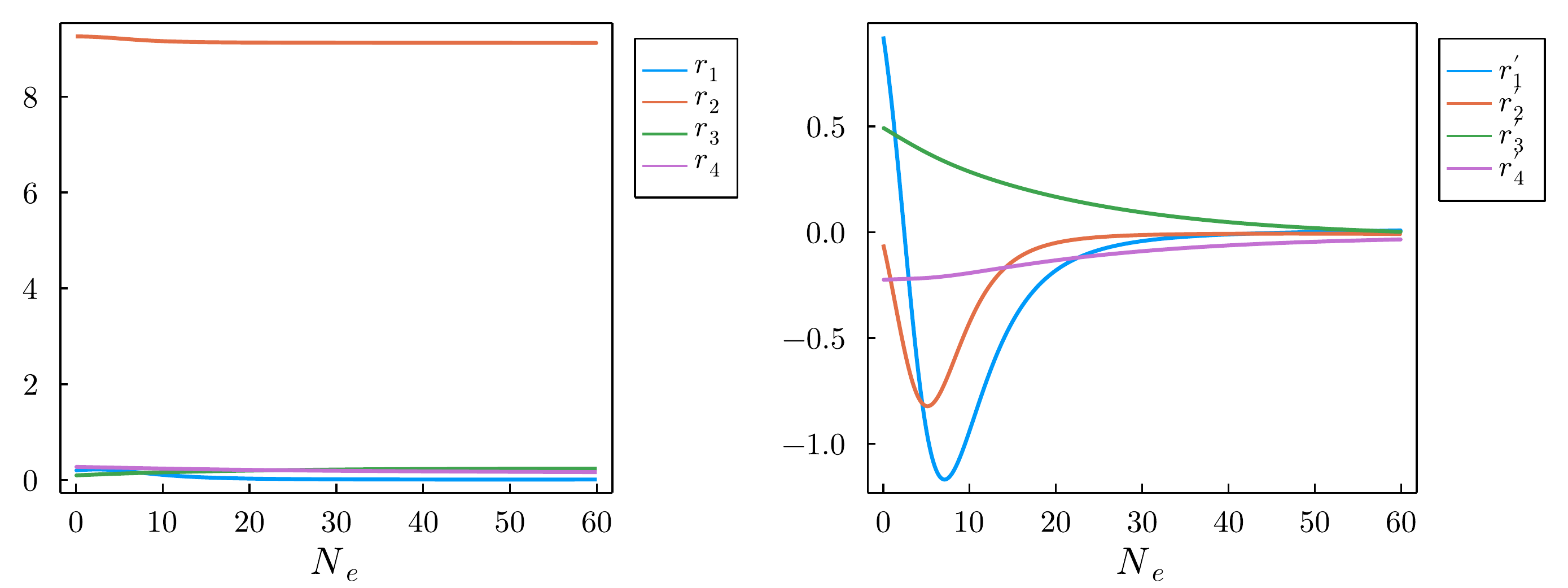}
\caption{A rapid-turn solution in the model \eqref{eq:poly1alphapotential}, \eqref{eq:poly1alphametric}. As expected from the metric structure, the majority of the field motion lies in the $(r_1,r_2)$ plane. The initial conditions chosen match those in Fig \ref{fig:poly1alpha}.}
\label{fig:poly1alphafields}
\end{figure}

\FloatBarrier
\section{Summary} \label{sec:summary} 

In this work we have sought to describe, in general, when multi-field inflationary trajectories lead to stable late-time attractor solutions.
When a late-time attractor exists, it should be seen as theoretically appealing since it avoids fine-tuning problems while allowing for compatibility with UV-inspired models which necessarily have many fields.
We focus on turning trajectories as they are uniquely multi-field, and are also phenomenologically interesting for their ability to source subhorizon growth of the curvature power spectrum, and potentially primordial black holes and/or gravitational waves.

In Section \ref{sec:frenet_serret} we describe inflationary trajectories in general and lay out our kinematic-basis framework for finding late-time attractor solutions.
In Section \ref{sec:two_field}, we recover known two-field rapid-turn attractors in our framework, while clarifying some 
notions on stability.
In Section \ref{sec:3field}, we extend this analysis to three fields, and characterize attractor solutions in the slow-twist limit.
We confirm several known three-field examples from the literature are well described by our slow-twist expressions, even when torsion is present ($T\lesssim\Omega$).
In Section \ref{sec:isometries}, we attempted a generalization of our procedure to an arbitrary number of fields under the assumption that the isometry structure of the field space metric is highly constrained.
As we later describe, several phenomenologically interesting models, from supergravity and elsewhere, fit the isometry structures we study.
These allow for a rapid identification of the allowed late-time attractor solutions from the form of the field space alone.
In Section \ref{sec:observables_stability}, we study inflationary perturbations with the goal of characterizing the observables of our attractor solutions.
We study explicit models and confirm our analytic understanding of the perturbations' superhorizon behavior. In subsequent works by the authors it was shown that slow- or rapid-twist inflation has observational consequences as it leads to distinctive features in the power spectrum; we refer the reader to \cite{Aragam:2023adu,Christodoulidis:2023eiw} for more details.
Lastly, as an application, we study in detail explicit many-field models from supergravity and elsewhere, and give the reader an explicit walk-through of the application of our methods.

This work shows the existence of a slow-twist multi-field inflationary attractor at $\mathcal{N}\geq 3$ in a wide variety of scenarios, derives general expressions for the inflationary dynamics on the attractor, and confirms them with analytic and numeric study of explicit models, including a few from supergravity.
These novel results will aid future multi-field model-building and analysis, and give us a deeper understanding of some of the most phenomenologically interesting inflationary scenarios.


\begin{acknowledgments}
RR would like to thank Sonia Paban and Vikas Aragam for helpful discussions during the course of this work and comments on an early draft of this manuscript. We would also like to thank the anonymous referee for suggestions which helped us in improving the clarity of this work. 
PC was supported in part by the National Research Foundation of Korea Grant 2019R1A2C2085023. RR was supported by an appointment to the NASA Postdoctoral Program at the NASA
Marshall Space Flight Center, administered by Oak Ridge Associated Universities under contract with NASA.
\end{acknowledgments}

\appendix

 \section{Linking the Frenet-Serret equations with the equations of motion}
 \label{sec:frenet_serret_app}

In this section we calculate the vectors of the orthonormal basis in terms of kinematic quantities ($\epsilon, \eta, \Omega, \cdots$) and dynamical quantities such as scalar products of derivatives of the potential. We start with the tangent vector $t^i$ and the potential gradient $w_i\equiv (\ln V)_{,i}$. We will first express these vectors in terms of covariant derivatives of the potential and then we will examine which are related with kinematic quantities. Using $\sigma'=\sqrt{2\epsilon}$ the normal vector is defined from 
\begin{equation} \label{eq:normal}
 k n^i \equiv \uD_{\sigma} t^i  = {3 - \epsilon \over 2\epsilon} (w_{\sigma}t^i - w^{i}) \, ,
\end{equation}
where $w_{\sigma} \equiv w_i t^i$ and $k \equiv \Omega / \sqrt{2\epsilon}$. Taking the projection along $n_i$ yields the following useful relation 
\begin{equation}
k = -  {3 - \epsilon \over 2\epsilon} w_n \, ,
\end{equation}
with $w_n\equiv w_i n^i$. Because of Eq.~\eqref{eq:normal} the gradient vector, written in the orthonormal basis, has only two non-vanishing components while $w_{\alpha}=0$ for $\alpha>2$. To find the next vectors in the series we will apply successively the derivative operator $\uD_{\sigma}$. We will need the following useful relations
\begin{align} \label{eq:wsigma1}
\uD_{\sigma} w_{\sigma} &= w_{\sigma \sigma} - { 2\epsilon k^2 \over 3 - \epsilon}    \, , \\
\uD^2_{\sigma} w_{\sigma} &= w_{\sigma \sigma \sigma} + 2k w_{\sigma n} + \epsilon_{,\sigma} w_n^2 - (3 - \epsilon) 2 w_n (w_{\sigma n}  - k w_{\sigma} ) \, .
\end{align}
Now, we apply the derivative operator on Eq.~\eqref{eq:normal} and using Eq.~\eqref{eq:wsigma1} we find the torsion vector as
\begin{equation} \label{eq:torsionapp}
\tau b^i =\left[ \left(\ln{3-\epsilon \over 2\epsilon k} \right)_{,\sigma } +  {3 - \epsilon \over 2\epsilon}  w_{\sigma}  \right] n^i  + {3 - \epsilon \over 2\epsilon k} w_{\sigma \sigma}t^i    -  {3 - \epsilon \over 2 \epsilon k}  w^{i}_{~;j}t^j  \, .
\end{equation}
The same equation shows that the rest $w_{\sigma b_i}$ components of the Hessian are zero. Applying once more the derivative operator we obtain
\begin{equation}
\begin{aligned} 
&\tau_2 b_2^i =\left[  \tau  + \left( - {1 \over \tau} \left(\ln { 2 \epsilon k \over 3 - \epsilon } \right)_{,\sigma } +  {3 - \epsilon \over 2\epsilon \tau}  w_{\sigma}  \right)_{,\sigma} +{(3-\epsilon) w_{\sigma \sigma} \over 2\epsilon \tau} \right] n^i + \left[ -\left(\ln { 2 \epsilon k \over 3 - \epsilon } \right)_{,\sigma } +  {3 - \epsilon \over 2\epsilon}  w_{\sigma}  \right] b^i  + \\
&  \left[ - {k \over \tau} \left(\ln { 2 \epsilon k \over 3 - \epsilon } \right)_{,\sigma } +  k{3 - \epsilon \over 2\epsilon \tau}  w_{\sigma}  + \left(w_{\sigma \sigma} {3 - \epsilon \over  2 \epsilon k \tau} \right)_{,\sigma}  \right]  t^i + \left(\ln  { 2 \epsilon k  \tau \over 3 - \epsilon }\right)_{,\sigma} {3 - \epsilon \over  2 \epsilon k \tau} p^{i}_{~;j}t^j  -  {3 - \epsilon \over  2 \epsilon k \tau } p^{i}_{~;jk}t^jt^k -  {3 - \epsilon \over 2\epsilon \tau}  p^{i}_{~;j}n^j \, ,
\end{aligned}
\end{equation}
and using again Eq.~\eqref{eq:torsion} we arrive at
\begin{equation}
\begin{aligned} 
&\tau_2 b_2^i = \left[   {3 - \epsilon \over 2\epsilon}  w_{\sigma} - \left(\ln \left( { 2 \epsilon k \over 3 - \epsilon }  \right)^2 \tau \right)_{,\sigma} \right] b^i  + \left[ w_{\sigma \sigma,\sigma} {3 - \epsilon \over  2 \epsilon k \tau}   - {k \over \tau} \left(\ln  { 2 \epsilon k \over 3 - \epsilon } \right)_{,\sigma } +  {k(3-\epsilon) \over 2\epsilon \tau}  w_{\sigma}  \right]  t^i   +    \\
&   \left[  \tau  + {1 \over \tau} \left( - \left(\ln  { 2 \epsilon k \over 3 - \epsilon } \right)_{,\sigma } +  {3 - \epsilon \over 2\epsilon \tau}  w_{\sigma}  \right)_{,\sigma} +{(3-\epsilon) w_{\sigma \sigma} \over 2\epsilon \tau} + {1 \over \tau} \left(\ln  { 2 \epsilon k \over 3 - \epsilon } \tau\right)_{,\sigma}  \left( -\left(\ln  { 2 \epsilon k \over 3 - \epsilon } \right)_{,\sigma } +  {3 - \epsilon \over 2\epsilon}  w_{\sigma}  \right)\right] n^i  \\
&-  {3 - \epsilon \over  2 \epsilon k \tau} p^{i}_{~;jk}t^jt^k -  {3 - \epsilon \over 2\epsilon \tau}  p^{i}_{~;j}n^j \, .
\end{aligned}
\end{equation}

These expressions relate projections of $w_{;ij\cdots}$ with the curvatures of the Frenet-Serret equations:
\begin{align}
k &= -{3- \epsilon \over 2\epsilon } w_i n^i \, , \\
\label{eq:torsionNormFS}
\tau &= - {3- \epsilon \over 2\epsilon k} w_{i;j}b^i t^j  \, , \\ 
\tau_2 &= - {3- \epsilon \over 2\epsilon k \tau} w_{i;jm} b_2^i t^j t^m  -  {3 - \epsilon \over 2\epsilon \tau}  w_{i;j}n^jb_2^i\, ,  
\end{align}
and so kinematic quantities are related to only specific components of the Hessian.

\section{Killing vectors of the hyperbolic space in two dimensions} \label{app:hyper}
Solving the Killing equation for the hyperbolic space in the exp representation \eqref{eq:hyperinflation} gives us the following generic form of the vector depending on three constants
\begin{equation}
\xi^i = \left( c_3 \chi + c_1, -{1 \over L} \chi \left( {c_3 \over 2} \chi + c_1 \right) + {L \over 2} c_3 e^{-2 \phi /L}+ c_2 \right) \, .
\end{equation}
Setting each time two out of the three constants to zero gives us the following three linearly independent Killing vector fields
\begin{align}
K_1^i &= (0,1) \, , \\ 
K_2^i &= \left(1, -{1 \over L} \chi \right) \, , \\
K_3^i &= \left( \chi, -{1 \over 2L} \chi^2 + {L \over 2} e^{-2 \phi /L}  \right) \, .
\end{align}
The first vector is associated with shifts in $\chi$ whereas the second and third are associated with shifts in the other isometry directions which become manifest once the hyperbolic metric is written in the cosh and sinh parameterization respectively. The Killing vectors $K_2,K_3$ depend on $\chi$ and so will be the normalized velocities along these vectors, which should not be the case if we consider a symmetric potential, as in the hyperinflation scenario, because $\chi$ is a cyclic variable for these models. For a rotationally symmetric potential $V=V(\phi)$ the canonical momentum is conserved
\begin{equation}
J = a^3 e^{2\phi/L} \dot{\chi} \, .
\end{equation}
Casting the Friedman constraint into a more convenient form 
\begin{equation}
{\ud \over \ud t} \left( a^3H \right) = a^3 V \, ,
\end{equation}
we can write the evolution equation for the $\phi$ field as
\begin{equation}
{\ud \over \ud t} \left( {\dot{\phi} + p H \over e^{2\phi / L} \dot{\chi}} - {\chi \over L}\right) = {\ud \over \ud N} \left( {\phi' + p  \over e^{2\phi / L} \chi'} - {\chi \over L}\right) = - a^3 p' \, ,
\end{equation}
where $p \equiv V_{,\phi}/V$. If additionally the potential is an exponential, $V\propto e^{p\phi}$, or if $p$ is slowly varying then
one finds another (approximate) integral of motion. Solving for $\chi$ yields
\begin{equation}
\chi = L {\phi' + p  \over e^{2\phi / L} \chi'} + C  \, , 
\end{equation} 
and in combination with the asymptotic expression for $\epsilon$ from Eq.~\eqref{eq:asymp_epsilon}, $2\epsilon \approx - p \phi'$, we finally find that the normalized velocities and gradients of the potential  in the other two Killing and orthogonal directions become asymptotically
\begin{align}
( u_2, u_{m2} )&  \rightarrow  ( -\sqrt{2\epsilon} , 0)  \, , \qquad  &( w_2, w_{m2} ) \rightarrow  (\sqrt{2\epsilon} ,- \sqrt{p^2 -  2\epsilon}) \, , \\ \label{eq:killing_3}
( u_3, u_{m3} )&  \rightarrow  ( \mp \sqrt{2\epsilon-  (\phi')^2} ,\phi' )  \, , \qquad  &( w_3, w_{m3} )  \rightarrow  (\pm 2\sqrt{2\epsilon-  (\phi')^2}  , -p -2 \phi') \, .
\end{align}
Based on these expressions we can investigate whether inflation can proceed along the Killlng or the orthogonal directions. For the second Killing direction, the velocity is aligned with the isometry direction and when this solution exists it describes rapid turn (because inflation proceeds along the isometry). For the third vector, we observe that the velocity in the orthogonal direction aligns with the $\phi$ field and can not be set to zero because the $\phi$ field will always decrease to smaller values as it descends its potential. In this case one can only set the isometry field to zero and the resulting solution is gradient flow. 

Investigating the three isometry directions we found that if inflation proceeds along/orthogonal to $K^i_1$ or $K^i_3$ only gradient flow is possible, whereas if it proceeds along $K^i_2$ we find a rapid-turn solution. Since the hyperbolic solution has significant turn rate it must be identified with this Killing direction.

\section{Coordinate transformations and isometries} \label{sec:app_cc}
In this section we demonstrate how to set the cross-correlations of the isometry fields with the orthogonal fields to zero. First, we note that in the coordinate system where the isometries are manifest $\{\chi, K_A \}$ the metric takes the general form 
\begin{equation}
\ud s^2 = G_{\chi\chi}\ud \chi^2 + 2 G_{\chi A}  \ud \chi \ud K^A + G_{AB}\ud K^A \ud K^B \, .
\end{equation}
Now we redefine the isometry fields as follows
\begin{equation}
K^A = \phi^A + f^A(\chi) \, ,
\end{equation}
and so the metric becomes
\begin{equation}
\ud s^2 = \left( G_{\chi \chi}+ 2  f^A_{,\chi}  G_{\chi A} + G_{AB} f^A_{,\chi}f^B_{,\chi} \right) \ud \chi^2 + 2 \left( G_{\chi A}   + G_{AB}  f^B_{,\chi}\right) \ud \chi \ud\phi^A  + G_{AB}\ud\phi^A \ud\phi^B  \, .
\end{equation}
In order to set the $_{\chi A}$ components to zero the coordinate transformation needs to satisfy 
\begin{equation}
f^B_{,\chi} = - H_{AB} G_{\chi A}  \, ,
\end{equation}
where $H$ is the inverse of the truncated matrix $G_{AB}$  ($H\cdot G= I$). Finally, we can canonically normalize the orthogonal field and obtain the metric \eqref{eq:isometry_multi}.

\section{Eigenvalues of block matrices} \label{app:block}
In this section we will show explicitly how to find the eigenvalues of a  block $2\mathcal{N} \times 2\mathcal{N}$ matrix $J$ of the following form
\begin{equation} \label{eq:matrix_J}
J = 
\begin{pmatrix}
0 & -I \\
\mathcal{M} & d I
\end{pmatrix} \, ,
\end{equation}
where $I$ is the identity matrix, $d$ a constant and each submatrix has dimensions $\mathcal{N} \times \mathcal{N}$.  Using a theorem from algebra one can express the determinant of any matrix $\Lambda$ 
\begin{equation}
\Lambda = 
\begin{pmatrix}
A & B\\
C & D
\end{pmatrix} \, ,
\end{equation}
in terms of the determinant of each block
\begin{equation}
\text{det}(\Lambda)= \text{det}(A) \text{det}(D-C A^{-1}B)\, .
\end{equation}
Now we specialize $\Lambda$ to the $J-\lambda I$
\begin{equation}
\Lambda = 
\begin{pmatrix}
-\lambda I  & - I \\
\mathcal{M} & (d-\lambda) I
\end{pmatrix} \, ,
\end{equation}
and using the previous formula we obtain
\begin{equation}
\text{det}(\Lambda)= \text{det}(-\lambda I) \text{det}\left( (d-\lambda)I + {1\over -\lambda}\mathcal{M} \right) =  \text{det}\left( (\lambda^2 -d \lambda)I + \mathcal{M} \right) \, .
\end{equation}
Decomposing the matrix $\mathcal{M}$ as $\mathcal{M}=U^{-1} D_{\mathcal{M}} U$, where $U$ is the matrix that diagonalizes it and $D_{\mathcal{M}}$ denoting the diagonal matrix with entries its eigenvalues, we find the determinant as
\begin{equation}
\text{det}(J)= \text{det}\left( (\lambda^2 -d \lambda)I + D_{\mathcal{M}} \right) \, .
\end{equation}
Setting the determinant to zero we find the eigenvalues $\lambda$ as
\begin{align}
\lambda_{a,\pm} = {1 \over 2} \left( d \pm \sqrt{d^2 - 4 \mu_{a}}\right) \, ,
\end{align}
where $\mu_a$ denotes the eigenvalues of $\mathcal{M}$. If the dynamical system of interest is of the form $X' = - J \cdot X$ then the condition for stability is $\text{Re}(\lambda_a)>0$, thus imposing restrictions on the eigenvalues of $\mathcal{M}$. Unless the matrix $\mathcal{M}$ is symmetric its eigenvalues are in general complex, and a complex eigenvalue $\mu$ with an imaginary part exceeding certain values will introduce a positive contribution to the real part of $\lambda$  which can potentially ruin stability. Recall that the real part of the square root of a complex number is given by the following formula
\begin{equation}
\text{Re}\left( \sqrt{x+i y} \right) = \pm  {1 \over \sqrt{2}} \sqrt{x + \sqrt{x^2 + y^2}} \, .
\end{equation}
Applying this formula for a complex eigenvalue $\mu$ we find the real part of $\lambda$ as
\begin{equation}
\text{Re}\left(\lambda_{a} \right) =  {1 \over 2} \left( d \pm {1 \over \sqrt{2}} \sqrt{d^2 - 4 \text{Re}\left(\mu_{a} \right) + \sqrt{\left[ d^2 - 4 \text{Re}\left( \mu_{a} \right) \right]^2 + 16 \text{Im}(\mu_{a})^2}}\right) \, .
\end{equation}
The previous set of eigenvalues have non-positive real part given that the real and imaginary parts of $\mu$ satisfy
\begin{align}
\text{Im}(\mu_{a})^2 < d^2\text{Re}(\mu_{a}) \, ,
\end{align}
which summarizes the conditions for stability.

The matrix \eqref{eq:matrix_J} with $d=3-\epsilon$ appeared in the linearized analysis of Secs.~\ref{subsec:n_field_stability} and \ref{subsec:supergravity}. When the torsion is zero and $\mathcal{M}=\mathcal{M}_{\rm sh}$ then this is also the stability matrix of orthogonal perturbations. For three fields the eigenvalues of $\mathcal{M}_{\rm sh}$ are
\begin{align}
\mu_\pm = {1 \over 2} \left( \mathcal{M}_{bb} +  \mathcal{M}_{nn} + 3 \Omega^2 \pm \sqrt{ \left( \mathcal{M}_{bb} +  \mathcal{M}_{nn} +3 \Omega^2\right)^2 - 4 \left[ \mathcal{M}_{bb}  ( \mathcal{M}_{nn} +3 \Omega^2) -  \mathcal{M}_{nb}^2 \right] }  \right) \, ,
\end{align}
or in a more compact form in terms of the trace and determinant  of $\mathcal{M}_{\rm sh}$
\begin{align} \label{eq:li}
\mu_\pm = {1 \over 2} \left( \text{tr}(\mathcal{M}_{\rm sh}) \pm \sqrt{  \text{tr}(\mathcal{M}_{\rm sh})^2 - 4 \text{det}(\mathcal{M}_{\rm sh} ) }  \right) \, .
\end{align}
The two eigenvalues become complex when the radical of Eq.~\eqref{eq:li} is negative, in which case the imaginary part is equal to 
\begin{align}
{1 \over 2}  \sqrt{ - \text{tr}(\mathcal{M}_{\rm sh})^2 + 4 \text{det}(\mathcal{M}_{\rm sh} ) }  \, ,
\end{align}
and we find the following condition for stability 
\begin{align} \label{eq:condition_third}
0< 4 \text{det}(\mathcal{M}_{\rm sh} ) < 2 (3 - \epsilon)^2  \text{tr}(\mathcal{M}_{\rm sh})  +  \text{tr}(\mathcal{M}_{\rm sh})^2  \, .
\end{align}
These are exactly the conditions \eqref{eq:superhorizon_criteria} for $T=0$.

\section{Numerical solution for powerspectra}
\label{sec:numerics}

To compute the powerspectra shown in this work, we use the transport method as first described in \cite{Dias:2015rca}, implemented in \texttt{Inflation.jl} \cite{inflationjl}, an open-source Julia-language multi-field inflation simulation package. For brevity, we do not elaborate on the method here, but note that it solves for the two-point function of the perturbations directly, rather than solving the system as written in Sec \ref{subsec:three_field_perturbations} and is widely regarded to be numerically stable.
For the results presented here, we begin the evolution by imposing Bunch-Davies initial conditions 8 e-folds before horizon exit of each mode, and solve for 5 modes around the pivot scale, equally log-spaced in the 5 e-folds centered at $k_\star$. We display the pivot scale mode's evolution as a function of $N$, and use the dependence of $P_\zeta$ on the 5 $k$-values to estimate the spectral index $n_s$.
We project the matrix of two-point functions along the kinetic basis vectors to construct the adiabatic $P_\zeta(k,N)$ and the isocurvature $P_\mathrm{iso}(k,N) \equiv P_S(k,N) + P_b(k,N)$ powerspectra. The powerspectrum of primordial tensor modes is also computed by \texttt{Inflation.jl} through another set of transport equations, and is displayed as the appropriately comparable amplitude to the scalar powerspectra, i.e. $P_\mathrm{tens}(k,N) = r(k,N) P_\zeta(k,N)$, where $r$ is the canonical scalar-to-tensor ratio.

\section{Initial condition search strategy}
\label{sec:blackboxing}

For the high-dimensional parameter space in section \ref{sec:manyfieldssim}, a manual search of initial conditions and parameter values would be extremely tedious at low $\mathcal{N}$, and impossible at high $\mathcal{N}$.

We therefore find suitable initial conditions in code, using an efficient differential evolution optimizer\footnote{\texttt{BlackBoxOptim.jl} -- \url{https://github.com/robertfeldt/BlackBoxOptim.jl}}, applied to the open-source multi-field inflation simulation package \texttt{Inflation.jl} \cite{inflationjl}.
The optimizer varies the initial field values and parameters in order to minimize a cost function.
Each iteration, we simulate background evolution at each point probed in parameter space. These points are then ranked by their cost, chosen to be
\begin{align}
\mathrm{cost}\left(\vec{\phi}(t=0), \mathrm{params}\right) &\equiv \frac{1}{N_e} + A + B\left[ \frac{1}{\Omega_\mathrm{end}} +\eta_\mathrm{max} \right],
\end{align}
where $N_e$ is the total number of e-folds of inflation before $\epsilon = 1$ or 60 e-folds have elapsed, whichever occurs first; $A = 10^6$ if $N_e < 60$, and is otherwise zero; $B$ is zero when $N_e < 60$, and otherwise 1.
$\Omega_\mathrm{end}$ is the minimum value of $\Omega$ in the final 30 e-folds, and $\eta_\mathrm{max}$ is the maximum absolute value of $\eta \equiv \epsilon^\prime / \epsilon$ during the same period.

This cost function is constructed to prefer long-lived trajectories first, and then later in the optimization refine them into slow-roll, rapid turn trajectories.
In practice, we chose $\mathcal{O}(100)$ different random seeds for the optimizer, let each optimize for $\sim 10 \, \mathrm{ min}$, and took the lowest cost result from the entire set.

This procedure proved efficient and finding suitable initial conditions, even in $\mathcal{O}(50)$-dimensional parameter space.

\end{document}